\documentclass[a4paper,10pt]{article}
\usepackage[round]{natbib}
\usepackage{comment}

\usepackage{enumerate}
\usepackage{times,amsmath,graphicx}
\usepackage{color}
\usepackage{multirow}
\usepackage[flushleft]{threeparttable}
\usepackage{bm}
\usepackage{geometry}
\usepackage{nohyperref}
\usepackage{algorithmic}
\usepackage{amsfonts}
\usepackage{algorithm}
\usepackage{url}
\usepackage[symbol]{footmisc}

\usepackage{etoolbox}
\makeatletter
\patchcmd{\@verbatim}
  {\verbatim@font}
  {\verbatim@font\small}
  {}{}
\makeatother
\usepackage{listings}
\usepackage{titlesec}
\titleformat*{\section}{\large\bfseries}
\titleformat*{\subsection}{\large\bfseries}
\titleformat*{\subsubsection}{\normalsize\bfseries}

\newcommand{\f}{\operatorname}

\newtheorem{theorem}{Theorem}[section]

\newtheorem{example}[theorem]{Example}

\begin{document}
\title{\huge \textbf{Sampling with censored data: a practical guide}}

\author{
    Pedro L. Ramos$^{\rm a*}$,
    Daniel C. F. Guzman$^{\rm b,c}$,
    Alex L. Mota$^{\rm d}$, \\
    Daniel A. Saavedra$^{\rm a}$,
    Francisco A. Rodrigues$^{\rm b}$, and
    Francisco Louzada$^{\rm b}$
    \vspace{0.2cm} \\
    $^{a}${Faculty of Mathematics, Pontificia Universidad Católica de Chile, Santiago, Chile} \\
    $^{b}${Institute of Mathematics and Computer Science, University of São Paulo, São Carlos, Brazil} \\
    $^{c}${Department of Statistics, Federal University of São Carlos, São Carlos, Brazil} \\
    $^{d}${Department of Statistics, Federal University of Amazonas, Manaus, Brazil} \\
    $^{*}${corresponding author: pedro.ramos@uc.cl}
}
\date{}

\maketitle

\begin{abstract}
In this review, we present a simple guide for researchers to obtain pseudo-random samples with censored data. We focus our attention on the most common types of censored data, such as type I, type II, and random censoring. We discussed the necessary steps to sample pseudo-random values from long-term survival models where an additional cure fraction is informed. For illustrative purposes, these techniques are applied in the Weibull distribution. The algorithms and codes in R are presented, enabling the reproducibility of our study. Finally, we developed an R package that encapsulates these methodologies, providing researchers with practical tools for implementation.
 \vspace{0.2cm} 
\end{abstract}

\section{Introduction} \label{sec1}

The recent advances in data collection have provided a unique opportunity for scientific discoveries in medicine, engineering, physics, and biology~\citep{shilo2020axes, mehta2019high}. Although some applications offer a vast amount of data, as in astronomy or genetics, there are some other scientific areas where missing recordings are very common~\citep{tan2016introduction}.  In many situations, data collection is limited due to certain constraints, such as temporal or financial restrictions, which do not allow for acquiring a data set in its complete form. Mainly, missing values are very common in clinical analysis, where, for instance, some patients do not recover from a given disease during an experiment, representing the missing observations. 

Survival analysis, which concerns the modeling of time-to-event data, provides tools to handle missing data in experiments \citep{klein2006survival}. Different approaches to deal with partial information (or censored data) have been discussed in the survival analysis theory, where the results are mostly motivated by time-to-event outcomes, widely observed in applications, including studies in medicine and engineering \citep{klein2006survival}. Although most books and articles have focused on discussing the theory and inferential issues to deal with censored data \citep{bogaerts2017survival}, it is challenging to find comprehensive studies discussing how to obtain pseudo-random samples under different types of censoring, such as type I, type II, and random censoring.  Type II censoring is usually observed in reliability analysis where components are tested, and the experiment stops after a predetermined number of failures \citep{singh2005estimation}. Hence, the remaining components are considered as censored \citep{tse2000statistical,  balakrishnan2007point,  kim2011bayesian, guzman2020linear}.

On the other hand, type I censoring is widely applied in (but not limited to) clinical experiments where the study is ended after a fixed time, and the patients who did not experience the event of interest are considered as censored \citep{algarni2020bayesian}. Finally, the random censoring consists of studies where the subjects can be censored during any experiment period, with different times of censoring \citep{qin2001empirical,  wang2001inference, wang2002empirical, li2003non, ghosh2019robust}. Random censoring is the most commonly used in scientific investigations since it is a generalization of types I and II censoring.

This paper provides a review of the most common approaches to simulate pseudo-random samples with censored data. A comprehensive discussion about the inverse transformation method, procedures to simulate mixture models, and the Metropolis-Hastings (MH) algorithm is presented. Furthermore, we provide the algorithms to introduce artificial censoring, including types I, II, and random censoring, in pseudo-random samples. The approach is illustrated in the Weibull distribution~\citep{weibull1939}, and the codes are present in language R, such that interested readers can reproduce our analysis in other distributions. Additionally, we also present an algorithm that can be used to generate pseudo-random samples with random censoring and the presence of long-term survival  \citep{rodrigues2009unification}. This algorithm will be useful to model medical studies where a part of the population is not susceptible to the event of interest \citep{chen1999new,  peng2000nonparametric, chen2002bayesian,  yin2005cure, kim2008cure}. 

To further enhance our contribution, we developed a new R package that specifically addresses the generation of pseudo-random samples with censored data. While several existing R packages provide similar functionalities for generating pseudo-random values for the Weibull and power-law distributions, such as the stats package (function rweibull; R Core Team, \citeyear{rcore2024}), the ExtDist package (function rWeibull; Wu et al., \citeyear{wu2023extdist}), the distributions3 package (function random.Weibull; Hayes et al., \citeyear{hayes2022distributions3}), the mixR package (function rmixweibull; Yu, \citeyear{yu2021finite}), and the poweRlaw package \citep{gillespie2015fitting}, our package offers additional functionalities specifically designed for handling censored data. It provides an integrated and user-friendly approach to analyzing censored data, making it a valuable addition to the existing suite of tools available to researchers.

This paper is organized as follows. Section 2 discusses the main properties of the Weibull distribution, which is the most well-known lifetime distribution. Section 3 describes the most common approaches to generate pseudo-random samples. Section 4 presents a step-by-step guide to simulate censored data under different censoring mechanisms and cure fractions. Section $5$  presents an extensive simulation study using different metrics to validate the results under many samples, illustrating the censoring effect on bias and the estimator's mean square error. Finally, Section 6 summarizes the study and discusses some possible further investigations.

\section{Weibull distribution}

Initially, we describe the Weibull distribution, which will be considered in the derivation of proposed sampling schemes. Let $X$ be a positive random variable with a Weibull distribution~\citep{weibull1939}, then its probability density function (PDF) is given by:
\begin{equation}\label{Eq:Weibull}
    f(x;\beta,\alpha)= \alpha\beta x^{\alpha-1} \exp\left\{-\beta x^\alpha\right\}, \quad x>0,
\end{equation} 
where $\alpha>0$ and $\beta>0$ are the shape and scale parameters, respectively. Hereafter, we will use $X\sim \text{Weibull}(\beta,\alpha)$ to denote that the random variable $X$ follows this distribution. The corresponding survival and hazard rate functions of the Weibull distribution are given, respectively, by
\begin{eqnarray}
S(x;\beta,\alpha)=\exp\left\{-\beta x^\alpha\right\}, \quad  \mbox{and}
\quad 
h(x;\beta,\alpha)=\dfrac{f(x;\beta,\alpha)}{S(x;\beta,\alpha)}=\alpha\beta x^{\alpha-1}.
\end{eqnarray}

The survival function of the Weibull distribution is proper, that is, $S(0)=1$ and $S(\infty)=\lim_{x \to \infty}S(x)=0$. Additionally, its hazard rate is always monotone with constant ($\alpha=1$), increasing ($\alpha>1$) or decreasing ($\alpha<1$) shapes. 

The Weibull distribution has two parameters, which are usually  unknown when dealing with a sample. The maximum likelihood approach is the most common method used to estimate the unknown parameters of the model. In this case, we have to construct the likelihood function defined as the probability of observing the vector of observations $\boldsymbol{x}$ based on the parameter vector $\boldsymbol{\theta}$. 
We obtain the maximum likelihood estimator (MLE) by maximizing the likelihood function. The likelihood function for $\boldsymbol\theta$ can be written as
\begin{equation*} L(\boldsymbol{\theta;x, \delta})=\prod_{i=1}^n f(x_i|\boldsymbol{\theta})^{\delta_i}S(x_i|\boldsymbol{\theta})^{1-\delta_i},\end{equation*}
where $\delta_i$ is the indicator of censoring in the samples with $\delta=1$ if the data is not censored and $\delta=0$ otherwise. Note that when $\delta_i=1, i=1,\ldots,n$ the likelihood function reduces to $L(\boldsymbol{\theta,x})=\prod_{i=1}^n f(x_i|\boldsymbol{\theta})$. The evaluation of the partial derivatives of the equation above is not an easy task. To simplify the calculations, the natural logarithm is considered because the maximum value at log-likelihood occurs at the same point as the original likelihood function. Notice that the natural logarithm is a monotonically increasing function. We consider the $\log$-likelihood function that is defined as  $\ell(\boldsymbol{x,\delta;\theta}) = \log L(\boldsymbol{\theta;x,\delta})$. Furthermore, the MLE, defined by $\hat{\boldsymbol{\theta}}(\boldsymbol{x})$, is such that,
 for every $x$, $ \widehat{\boldsymbol{\theta}}(\boldsymbol{x})
  = \arg\max_{\boldsymbol{\theta} \in \boldsymbol{\Theta}}\ell(\boldsymbol{x,\delta;\theta})$. This method is widely used due to its good properties, under regular conditions. The MLEs are consistent, efficient and asymptotically normally distributed, (see \citep{migon2014statistical} for a detailed discussion),
\begin{equation*} \boldsymbol{\hat{\theta}} \sim \text{N}_k\left(\boldsymbol{\theta},I^{-1}(\boldsymbol{\theta})\right) \quad \mbox{for} \quad n \to \infty , \end{equation*}
where $I(\boldsymbol{\theta})$ is the $k \times k$ Fisher information matrix for $\boldsymbol{\theta}$, where  $I_{ij}(\boldsymbol{\theta})$ is the $(i,j)$-th element of $I(\boldsymbol{\theta})$ given by
\begin{equation*}\label{fisherinf}
I_{ij}(\boldsymbol{\theta})=\text{E}\left[-\frac{\partial^2}{\partial \theta_i \partial \theta_j}\ell(\boldsymbol{\theta;x, \delta})\right]. 
\end{equation*}

In our specific case due to the presence of censored observations (the censorship is random and non-informative), we cannot calculate the Fisher information matrix  $I(\boldsymbol{\theta}).$ Therefore, an alternative approach is to use the observed information matrix $H(\boldsymbol{\theta})$ evaluated at $\hat{\boldsymbol{\theta}}$, i.e. $H(\hat{\boldsymbol{\theta}})$. The terms of this matrix are 
\begin{equation*}
H_{ij}(\hat{\boldsymbol{\theta}})=-\frac{\partial^2}{\partial \theta_i \partial \theta_j}\ell(\boldsymbol{\theta;x, \delta}) \bigg{|}_{\boldsymbol{\theta}=\hat{\boldsymbol{\theta}}}.
\end{equation*}

Large sample confidence intervals (approximate) at level $100(1-\xi)\%$, for each parameter $\theta_i$, $i=1,\,\ldots,\,k$, can be calculated as
\begin{equation*} 
\hat{\theta}_i \pm Z_{\frac{\xi}{2}}\sqrt{H^{-1}_{ii}(\hat{\boldsymbol{\theta}})},
\end{equation*}
where $Z_{\xi/2}$ denotes the $\left(\xi/2\right)$-th quantile of a standard normal distribution. The derivation of MLEs for each of the simulation schemes is presented in Section \ref{mleesqsimula}.

To perform the simulations, we consider the R language \citep{rcore2024}. This language provides a free environment for statistical computing, which renders many applications, such as machine learning, text mining, and graphs. The R language can also be compiled and run on various platforms, widely accepted and used in the academic community and industry. For high-quality materials for R practitioners, see \citep{robert2010introducing, suess2010introduction,field2012discovering,james2013introduction,schumacker2013understanding}.

\section{Sampling from random values}

In this section, we revise two standard procedures to sample from a target distribution. The procedures are simple and intuitive and can be used for most of the probability density functions introduced in the literature.

\subsection{The inverse transformation method}

The currently most popular and well-known method used for simulating samples from known distributions is the inverse transformation method (also known as inverse transform sampling)~\citep{ross2006first}).

The uniform distribution in $(0,1)$  plays a central role during the simulation procedures as, in general, the procedures are based on this distribution. There are many methods for simulating pseudo-random samples from uniform distributions. The congruential method and the feedback shift register method are the most popular ones, and other procedures are combinations of these algorithms \citep{niederreiter1992random}. Here, we consider the standard procedure implemented in R to generate samples from the uniform distribution.

Let $u$ be a realization of a random variable $U$ with distribution $U(0,1)$. From the  inverse transformation method, we have that $X=F^{-1}(u)$, where $F(.)$ is  a invertible distribution function, so that $F_X^{-1}(u) = \inf\{x;F(x)\geq u, u\in (0,1)\}.$ Hence, the method involves the computation of the quantile function of the target distribution.

The \textit{algorithm} describing this methodology takes the following steps:
\begin{algorithm}[H]
\caption{Inverse transformed method}
\label{wsimulinvertransf}
\begin{algorithmic}[1]
\STATE \mbox{Define} $n$, and $\boldsymbol{\theta}$, 
\FOR{$i=1$ to $n$}
\STATE $u_i \leftarrow U(0,1)$
\STATE $x_i \leftarrow F^{-1}(u_i,\boldsymbol\theta)$
\ENDFOR
\end{algorithmic}
\end{algorithm}

To illustrate the above procedure's applicability, we employ such a procedure in the Weibull distribution to generate pseudo-random samples. The calculations are performed using language R.  

\begin{example}Let $X\sim \text{Weibull}(\beta,\alpha)$. To apply the inverse transform method above to generate the pseudo-random samples, we consider the cumulative density function given by
\begin{equation*}
F(x;\beta,\alpha)=1-\exp\left\{-\beta x^{\alpha}\right\}, \quad x>0,
\end{equation*}
and after some algebraic manipulations, the quantile function is given by
\begin{equation}\label{random_inv}
   x_u=\left(\frac{\left(-\log(1-u)\right)}{\beta}\right)^{\frac{1}{\alpha}} .
\end{equation}

Using the R software, we can easily generate the pseudo-random sample by considering the following algorithm:
\end{example}

\begin{lstlisting}[language=R]
n=10000;  beta <- 2.5; alpha <- 1.5 #Define the parameters
rweibull<-function(n,alpha,beta) {
	U<-runif(n,0,1)
	t<-((-log(1-U)/beta)^(1/alpha))
	return(t)
}
x<-rweibull(n, alpha, beta)
\end{lstlisting}

\begin{figure}[!h]
	\centering
	\includegraphics[width=0.8\textwidth]{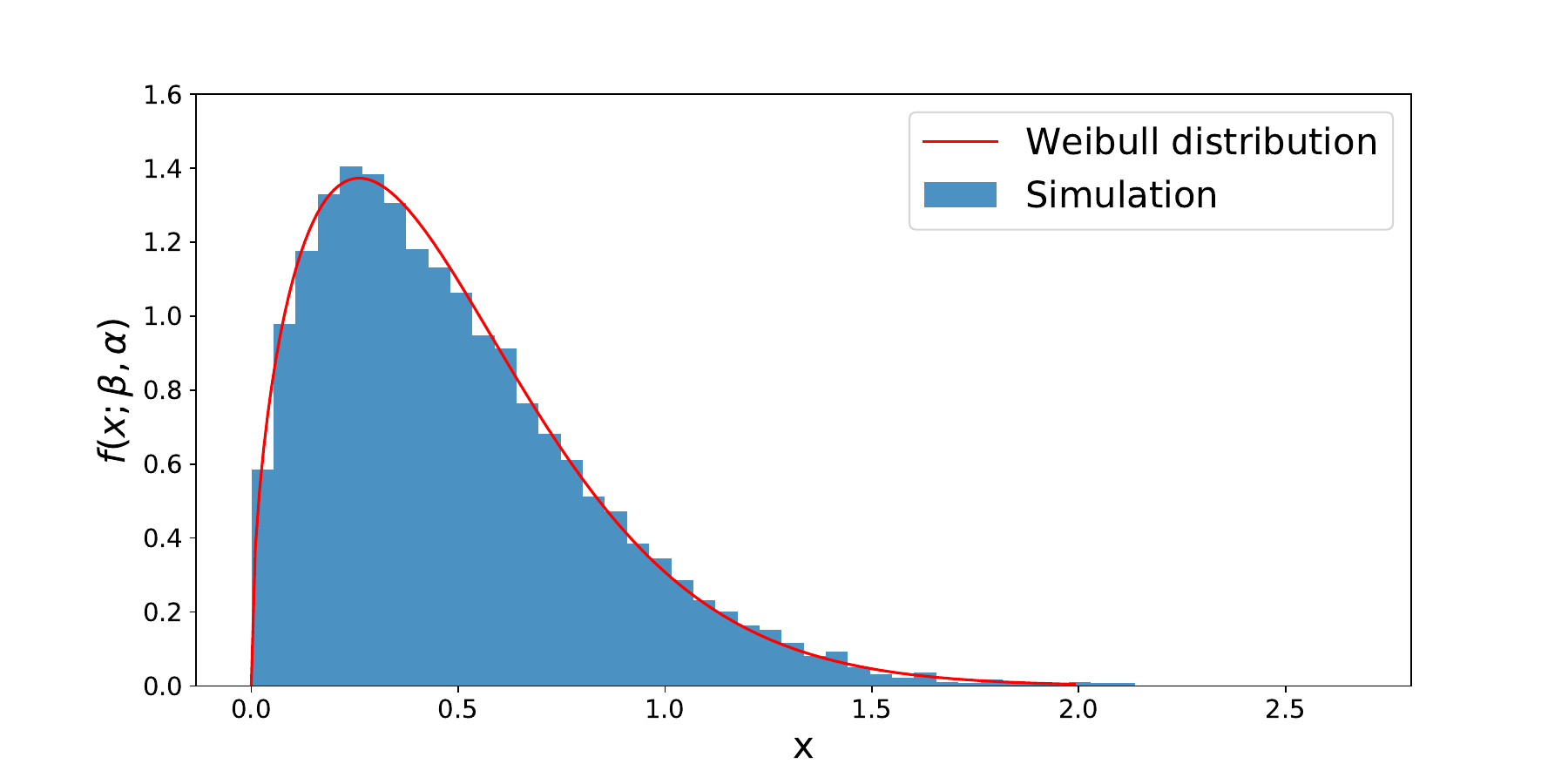}
	\caption{Histogram of 10,000 samples generated by the inverse transformation method applied in the Weibull distribution. The red line shows the theoretical probability density function shown in equation~(\ref{Eq:Weibull}). }\label{fig:weibull}
\end{figure}

The histogram of the obtained samples is shown in Figure~\ref{fig:weibull}.

The function above will be used in all the programs discussed throughout this paper. Hence, it should be used before running the codes. \textcolor{black}{It is important to note that functions for generating Weibull values are already available in R packages. Refer to \texttt{rweibull} in the \texttt{stats} package \citep{rcore2024} and \texttt{rWeibull} in the \texttt{ExtDis} package \citep{wu2023extdist}.
}

\subsection{Random samples for mixture models}

Mixture models have been the subject of intense research due to their wide applicability and ability to describe various phenomena whose analysis and adjustment would be almost impossible to perform with the assumption of a single distribution~\citep{dellaportas1997bayesian}. The mixture models were designed to describe populations comprising sub-populations where each one has its characteristic distribution. The mixture model's shape is a sum of each of the $ k $ components called the mixture component, each one multiplied by its corresponding proportion within the population. The mixing model is expressed as follows:
\begin{equation}\label{modelmistufi}
    f(x)=\sum_{j=1}^{k}p_{j}f(x|\mbox{\boldmath$\theta$} _{j}),
\end{equation} 
where the probabilities $p_j$, $j=1, \ldots, k$ with $ \sum_{i = 1}^{k}p_i = 1, $ corresponds to the proportion of each component of the population. The probability density $f(x)$, expressed in equation (\ref{modelmistufi}) with the vector $\mbox{\boldmath$ \theta$}_j$ of parameters, referring to the $j-$th component of the mixture, is said to have a $k$-finite mixture density. 

For the case of $k$ mixture components, we have that $ f_1(X), \ldots, f_k(X) $ are the $k$ densities of each factor with $ p_1, \ldots, p_k$ proportions, respectively. The algorithm increasingly sorts these proportions to generate samples from $f(x)$, so that the sample origin of the respective component can be identified, supposing that $ p_1 <\ldots <p_n $. We now describe the algorithm to generate samples from a mixture model.

\begin{algorithm}[H]
\caption{Simulation of random samples for mixture models}
\label{wsimulmixcensorship}
\begin{algorithmic}[1]
\STATE \mbox{Define} $n,k$, and $\boldsymbol{\theta}_j $, $p_j$, $j=1,\ldots k$, 
\STATE $t$ vector
\STATE $p_0=0$
    \FOR{$i=1$ to $n$}
    \STATE $u_i \leftarrow U(0,1)$
    \FOR{$j=1$ to $k$}
\IF{$(p_{j-1} < u_i \leq p_{j})$}
\STATE $t_i \leftarrow F_j ^{-1}(u_i,\boldsymbol{\theta}_j )$
\ENDIF
\ENDFOR
\ENDFOR
\PRINT t
\end{algorithmic}
\end{algorithm}

The algorithm presented above is straightforward to be applied. For instance, to generate data for mixture models with two components $k=2$, where each component is a  Weibull, we have the following code in R.

\begin{figure}[!h]
	\centering
	\includegraphics[width=0.8\textwidth]{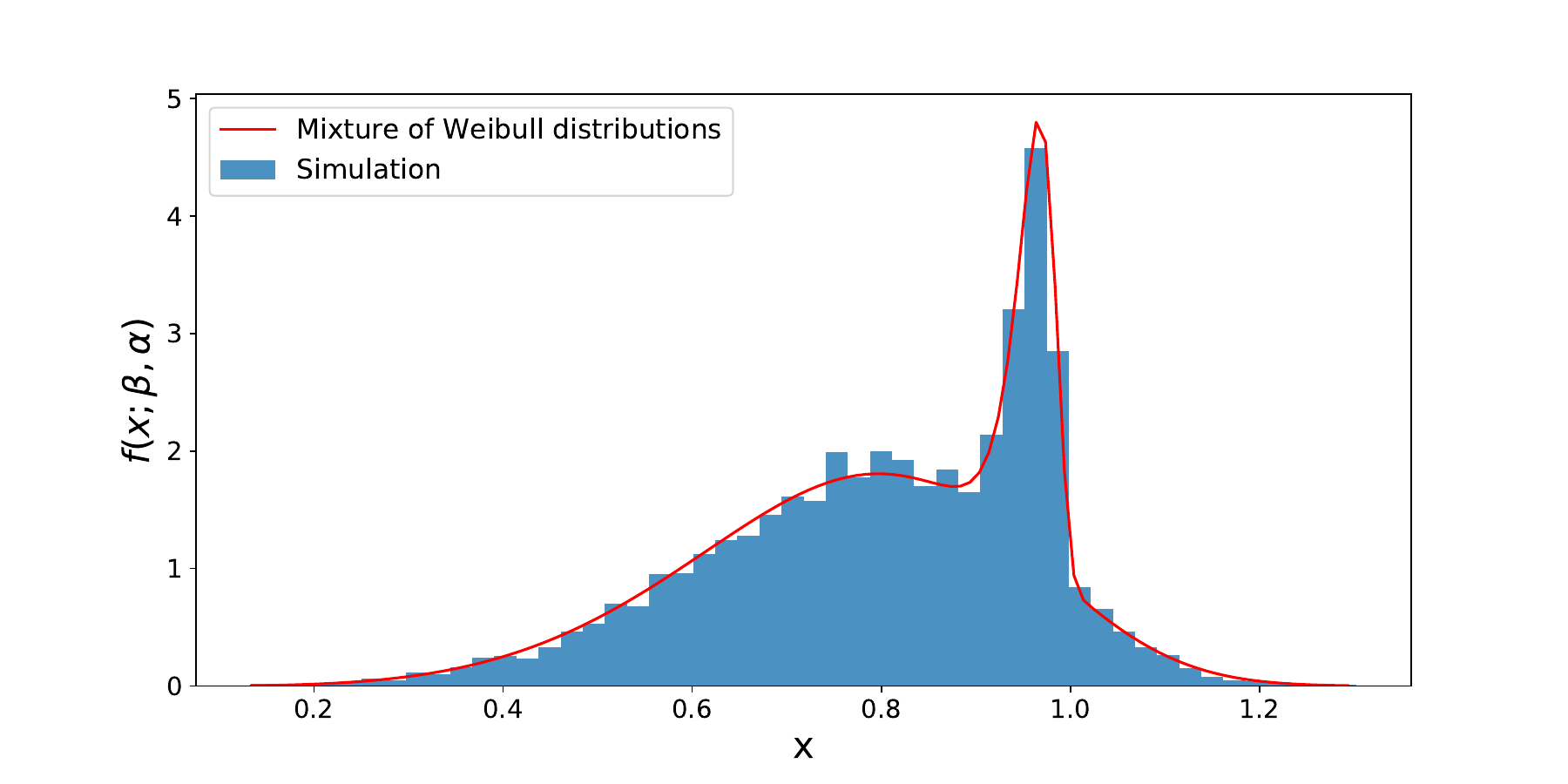}
	\caption{Histogram of 10,000 samples from a mixture of two Weibull distributions, generated by the inverse transformation method.}\label{fig:mixture}
\end{figure}

\begin{example} Let $X_1\sim \text{Weibull}(\beta_1,\alpha_1)$ and $X_2\sim \text{Weibull}(\beta_2,\alpha_2)$ with a probabilty density function given by
\begin{equation*}
f(x|\boldsymbol{\alpha},\boldsymbol{\beta})=pf_1(x|\beta_1,\alpha_1)+(1-p)f_2(x|\beta_2,\alpha_2),
\end{equation*}
where $f_j(x|\beta_j,\alpha_j)$ has density $\text{Weibull}(\beta_j,\alpha_j)$ for $j=1,2$ and $p$ is the proportion related to the first subgroup. The R codes to generate such a sample, given the parameters, are given as follows. In Figure~\ref{fig:mixture}, we show the simulation results.

\end{example}
\newpage

\begin{lstlisting}[language=R]
alpha1=5; 
beta1=2.5
alpha2=50; 
beta2=5
n=10000; 
p1=0.8; 
p2=1-p1
t<-c()
x1 <- rweibull(n,alpha1,beta1)
x2 <- rweibull(n,alpha2,beta2)
u<-runif(n,0,1)
for (i in 1:n) {
  if (u[i]<=p1){
    t[i] <- x1[i]
  }
  else{
    t[i] <- x2[i]
  }
}
\end{lstlisting}

\textcolor{black}{ It is worth mentioning that functions for generating a mixture of Weibull distributions have already been implemented in R and can be seen in \cite{yu2021finite}.}

\subsection{When the inverse transformation method fails}

The Markov chain Monte Carlo (MCMC) methods arise from simulating complex distributions where the inverse transformation method fails. These methods initially appeared in physics applications to simulate the behavior of particles and later found applications in other branches of science, including statistics. This subsection describes the Metropolis-Hastings (MH) algorithm, one of the MCMC methods. It also shows how it can be used to generate random samples from a given probability distribution.

The MH algorithm was proposed in 1953 \citep{metropolis1953equation} and was further generalized \citep{hastings1970monte}, with the Gibbs sampler \citep{gelman1992inference} as a particular case. This algorithm has played a fundamental role in Bayesian statistics \citep{chib1995understanding}. These works applied the approach in the simulation of both absolutely continuous target densities and mixed distributions of discrete and continuous types. Here, we provide a brief review of the concepts needed to understand the MH algorithm. This presentation is not exhaustive, and a detailed discussion of MCMC methods is available in \cite{gamerman2006markov}, \cite{robert2010introducing}, and the references therein.

An important characteristic of the algorithm is that we can draw samples from any probability distribution $F(\cdot)$ if we know the form of $f(x)$, even without its normalized constant. Therefore, this algorithm is handy for generating pseudo-random samples even when the probability density function (PDF) has a complex form. To obtain this sample, we need to consider a proposal distribution $\mathrm{q}(\cdot)$. This proposed distribution also depends on the values that were generated in the previous state, $x_{n-1}$, that is, $q(\cdot|x_{n-1})$. The initial value used to initiate the algorithm plays an important role as the Markov chain always converges to the same equilibrium distribution. However, note that the Markov chain must fulfill certain properties to ensure convergence to an equilibrium distribution, such as being irreducible and reversible \citep{hoel1986introduction}.


The Metropolis-Hastings algorithm is summarized next.

\begin{algorithm}
\caption{Metropolis-Hastings algorithm}
\label{MHalgorithm}
\begin{algorithmic}[1]

\STATE Start with an initial value $x^{(1)}$ and set the iteration counter $j=1 ;$

\STATE Generate a random value $x^{(*)}$ from the proposal distribution $Q(\cdot)$;

\STATE Evaluate the acceptance probability
\begin{equation}\label{aceptance}
\alpha\left(x^{(j)},x^{(*)}\right)=\min\left(1, \frac{f\left(x^{(*)}|\boldsymbol{x}\right)}{f\left(x^{(j)}|\boldsymbol{x}\right)} \frac{\f{q}\left(x^{(j)}|x^{(*)},b\right)}{\f{q}\left(x^{(*)}|x^{(j)},b\right)}\right),
\end{equation}
where $f(\cdot)$ is the probability distribution of interest. Generate a random value $u$ from an independent uniform in $(0,1)$;

\STATE If $\alpha\left(x^{(j)},x^{{(*)}}\right)\geq u(0,1)$ then $x^{(j+1)}=x^{(*)}$, otherwise $x^{(j+1)}=x^{(j)}$;

\STATE Change the counter from $j$ to $j + 1$ and return to step 2 until convergence
is reached.

\end{algorithmic}
\end{algorithm}

The proposal distribution should be selected carefully. In the Metropolis-Hastings algorithm, the acceptance probability is crucial for ensuring that the Markov chain converges to the target distribution $f(\cdot)$. The acceptance probability (\ref{aceptance})
is based on the ratio of the target distribution values, conditioned on $f(\cdot)$, and the proposal distribution values, conditioned on $q(\cdot)$. This conditioning ensures that the algorithm properly accounts for the relative likelihood of moving to the proposed state versus remaining in the current state.

The parameter $b$ in the proposal distribution $q(\cdot)$ influences the proposal's behavior, such as the step size or variance, affecting the efficiency and mixing properties of the Markov chain. Proper tuning of $b$ is essential for achieving good performance in the Metropolis-Hastings algorithm. For further details and a theoretical background on the Metropolis-Hastings algorithm, please refer to \cite{chib1995understanding}.

As an application, if we want to sample from the Weibull distribution where $x>0$, using the Beta distribution as $q(\cdot)$ with $0\leq x\leq 1$ is not a good choice as they have different domains. In this case, a common approach is to consider a truncated normal distribution with a range similar to the target distribution domain. Furthermore, the pseudo-random sample generated from the algorithm above can be considered as a sample of $f(\cdot)$. 

\begin{example} 
The Weibull distribution can be written as $f(x;\beta,\alpha)= C(\beta,\alpha)x^{\alpha-1} \exp\left\{-\beta x^\alpha\right\}$ where $C$ is a normalizing function. Suppose that we drop the power parameter $\alpha$ in $e^{-\beta x^\alpha}$, remove $x^{-1}$ and include the possibility of the presence of a lower bound for $x$, referred to as $x_{\min}$. Thus, we obtain 
\begin{equation}\label{pdfplc}
f(x\,| \alpha,\beta,x_{\min})=C(\alpha,\beta,x_{\min})x^{-\alpha}e^{-\beta x} ,
\end{equation}
where $\alpha>0$ and $\beta>0$ are the unknown shape and scale parameters, $C(\cdot)$ is the normalized constant given by
\end{example}
\begin{equation}
C(\alpha,\beta,x_{\min})=\frac{\beta^{1-\alpha}}{\Gamma(1-\alpha,\beta x_{\min})},
\end{equation}
where
\begin{equation}
\Gamma(s,x)=\int_{x}^{\infty}t^{s-1}e^{-t}dt,
\end{equation}
is the upper incomplete gamma function. This distribution is known as the power-law distribution with cutoff \citep{clauset2009power} and has a more complicated structure than its related power-law distribution. Note that in the limit when $\beta \rightarrow 0$, the cutoff term is 
\begin{equation*}
\lim_{\beta \to 0}=\frac{\beta^{1-\alpha}}{\Gamma(1-\alpha,\beta x_{\min})}=\frac{\alpha-1}{x_{\min}}\left(\frac{1}{x_{\min}} \right)^{-\alpha}.
\end{equation*}
From the expression above and as the exponential term tends to unity when $\beta \rightarrow 0$, we obtain the standard power-law distribution given by
\begin{equation}\label{pdfplc2}
f(x\,| \alpha,x_{\min})=\frac{\alpha-1}{x_{\min}}\left(\frac{x}{x_{\min}} \right)^{-\alpha}.
\end{equation}
The cumulative distribution function (CDF) of $f(x\,| \alpha,\beta,x_{\min})$ has the form,
\begin{equation*}\label{cdfplc}
\begin{aligned}
F(x\,| \alpha,\beta,x_{\min})&=\frac{\beta^{1-\alpha}}{\Gamma(1-\alpha,\beta x_{\min})}\int_{x_{\min}}^{x} y^{-\alpha}e^{-\beta y} dy.
\end{aligned}
\end{equation*}

Since the CDF does not have a closed-form expression, we cannot obtain the expression of the power-law model's quantile function in a closed form. While in most situations, we can obtain pseudo-random samples using a root finder with $u_i - F(x_i) = 0$, the obtained samples for the power-law distribution with cutoff are very different from the expected ones. The problem may occur due to the integrals involved in the CDF. To overcome this problem, we consider the Metropolis-Hastings algorithm. A function in R to sample from a power-law distribution can be seen as follows.
\newpage
\begin{lstlisting}[language=R]
library(ggdmc)
n<-10000; xm<-1; alpha<-1.5 ; beta<-0.5 ;
rplc<-function(n,xm,alpha,beta,burnin=1000,thin=50,se=0.5){
  R<-n*thin+burnin #Number of replication
  #Density without the normalized constant
  den <- function (x) {-alpha*log(x)-beta*x}
  xv<-length(R+1); 
  xv[1]<-xm+1;
  for(i in 1:R) {
    prop1<- rtnorm(1,xv[i],se,lower=xm, upper=Inf)
    dens1<-dtnorm(xv[i], prop1, se, lower=xm, upper=Inf, 
                  lg = TRUE)
    dens2<-dtnorm(prop1, xv[i], se,lower=xm, 
                  upper=Inf, lg = TRUE)
    ratio1<-den(prop1)-den(xv[i]) +dens1-dens2
    has<-min(1,exp(ratio1)); u1<-runif(1)
    if (u1<has & is.double(has)) {
      xv[i+1]<-prop1
    } 
    else {
      xv[i+1]<-	xv[i]
    }
  }
  x<-xv[seq(burnin,R,thin)]
  return(x)
}
x<-rplc(n,xm,alpha,beta) 

\end{lstlisting}

\begin{figure}[!h]
	\centering
\includegraphics[width=0.8\textwidth]{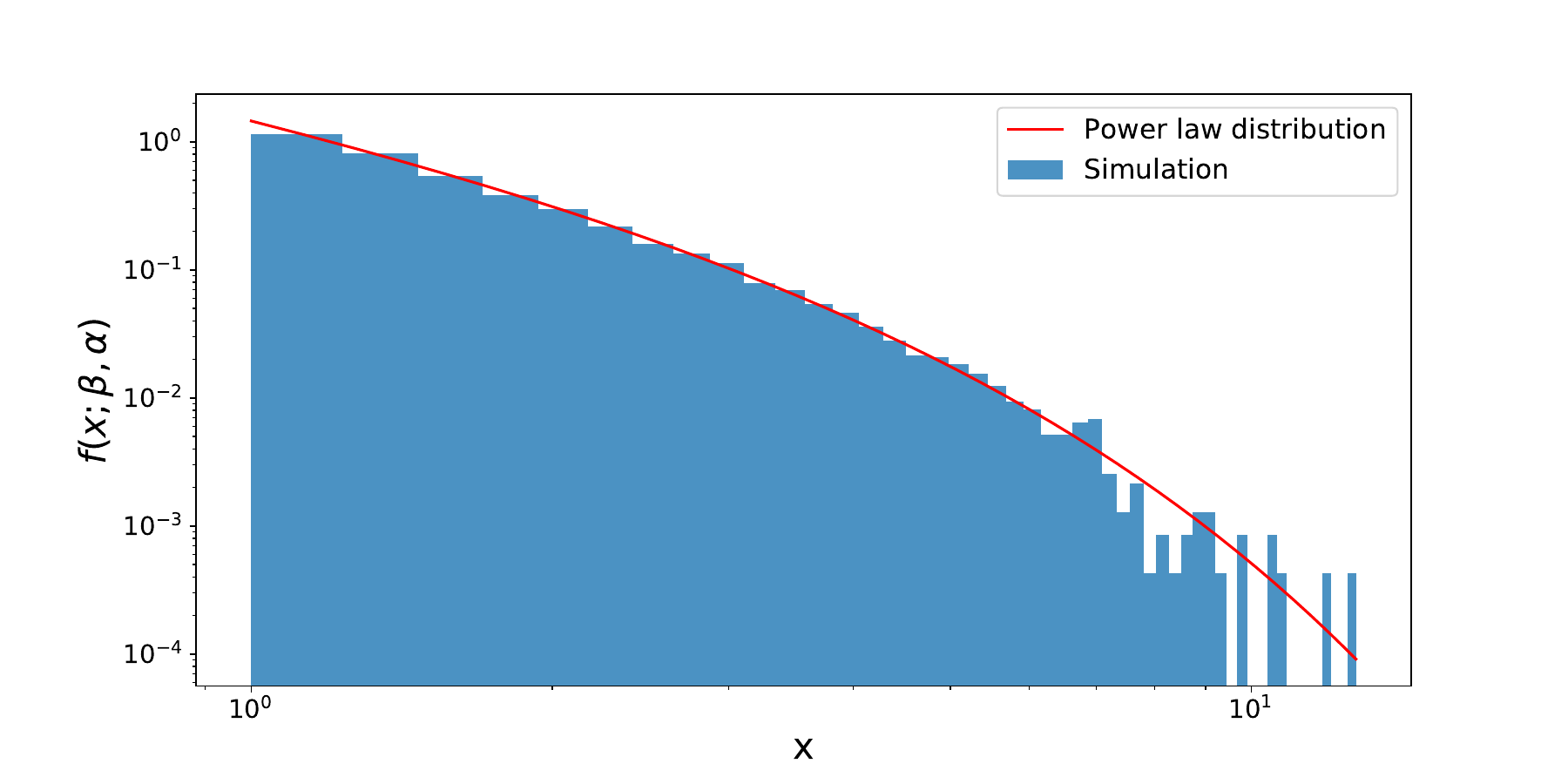}
	\caption{Histogram of 10,000 samples from the power-law distribution in equation~\eqref{pdfplc}, generated by the Metropolis-Hastings algorithm.}\label{fig:MH}
\end{figure}

The function \textit{rplc} can be used to obtain a pseudo-random sample from the power-law with a cutoff (PLC) distribution. In this case, some additional parameters need to be set: (i) the burning, (ii) the thin which is used to decrease the auto-correlation between the elements of the sample, and (iii) the \textit{se}, i.e., the standard error of the truncated normal distribution chosen as $q(\cdot)$. This distribution was assumed due to its flexibility to set a lower bound $x_{min}$, hence providing values in the same PLC distribution domain. For the additional parameters, we set default values for the burning $=1000$, thin $=50$, and se $=0.5$. Hence, if different values are not presented, the function will consider these values to generate the pseudo-random samples. Figure~\ref{fig:MH} shows the histogram of 10,000 samples generated by this function and the respective theoretical curve, given by equation~\eqref{pdfplc}.

To verify if the PLC parameters' estimates are closer to the true values, we derived the MLEs for the parameter of the PLC distribution, which can be seen in Appendix B. \textcolor{black}{Additionally, the estimation of the power-law distribution without a cutoff was discussed by \cite{gillespie2015fitting} and \cite{jerez2023scale}.}

\newpage 

\section{Sampling right-censored data }

Sampling pseudo-random data with censoring is usually a challenging task. In this section, we discuss simple procedures to generate such pseudo-random samples in the presence of type I, II, and random censoring. These censoring schemes are the most common in practice. In the end, we also present some references to generate other types of censoring.

\subsection{Type II censorship}

To generate the type II censorship, we define the total number of observations $n$ and the number $m$ of observations censored. 
First, we generate a sample of size $n$ according to a given distribution. We sort these observations in increasing order and associate the first $n-m$ values to the observed data. The remainder $m$ observations receive the value in position $m+1$. Since these observations did not experience the event of interest, we assume that their survival time is equal to the largest one observed.
The algorithm is described as follows.
\begin{algorithm}[H]
\caption{Simulation With Type II censorship}
\label{wsimultypeii2censorship}
\begin{algorithmic}[1]
\STATE \mbox{Define} $n$, $m$ and $\boldsymbol{\theta}$, 
\FOR{$i=1$ to $n$}
\STATE $u_i \leftarrow U(0,1)$
\STATE $x_i \leftarrow F^{-1}(u_i,\boldsymbol\theta)$
\STATE $\delta_i \leftarrow 1$
\ENDFOR
\STATE $t \leftarrow \text{sort}(x)$
\FOR{$i=n-m+1$ to $n$}
\STATE $t_i \leftarrow t_{n-m}$
\STATE $\delta_i \leftarrow 0$
\ENDFOR 
\PRINT $t$, $\delta$
\end{algorithmic}
\end{algorithm}

The algorithm described above is general and applicable to any probability distribution from which one can obtain a pseudo-random sample. It is important to note that the inverse transformation method is typically employed to acquire such a sample. However, this procedure can also be adapted for use with other sampling schemes. To illustrate the proposed approach, consider an example involving the Weibull distribution.

\begin{example}\label{exctype2}
Let $X\sim\text{Weibull}(\beta,\alpha)$. We can generate samples with type II censoring by fixing the number of elements to be censored. We will generate $n = 50$ observations from the Weibull distribution and set the number of censoring $m = 15$. The code in R is given as follows.

\end{example}
\begin{lstlisting}[language=R]
######Weibull R data simulation With Type II censorship#####
n=50  #Sample size
m=15  #Number of censored data
beta <- 2.5 ##Parameter value
alpha <- 1.5 ##Parameter value
x <- rweibull(n,alpha,beta)
t <- sort(x)
r<-n-m
t[(r+1):n]=t[r]
delta<-ifelse(seq(1:n)<(n-m+1),1,0) #Vector indication censoring
\end{lstlisting}

An example of the output obtained from the code above in R is given by:
\begin{verbatim}
> print(t)
 [1] 0.01 0.01 0.03 0.09 0.10 0.10 0.15 0.15 0.17 0.17 0.20 0.22 0.22 0.25 0.31
[16] 0.32 0.34 0.34 0.35 0.36 0.39 0.40 0.41 0.44 0.46 0.46 0.47 0.49 0.53 0.53
[31] 0.54 0.56 0.56 0.62 0.67 0.67 0.67 0.67 0.67 0.67 0.67 0.67 0.67 0.67 0.67
[46] 0.67 0.67 0.67 0.67 0.67
> print(delta)
 [1] 1 1 1 1 1 1 1 1 1 1 1 1 1 1 1 1 1 1 1 1 1 1 1 1 1 1 1 1 1 1 1 1 1 1 1 0 0 0
[39] 0 0 0 0 0 0 0 0 0 0 0 0

\end{verbatim}

We can see that 15 observations are censored and receive the largest lifetime among the observed lifetimes. The vector \emph{delta} stores the classification of each sample, where the value 0 is attributed to censored data.

\subsection{Type I censorship}

Type I censoring data is commonly observed in various applications related to medical and engineering studies. In this type of censoring, an experiment is terminated at a predetermined time, known as the censoring time. This is in contrast to type II censoring, where the study ends only after ``$n-m$" failures. Consider, for instance, a scenario where the reliability of a component is tested in a controlled study. Given the high reliability of many components, such an experiment could take several years to complete, potentially rendering the product outdated by its conclusion. To address this issue, the final time of the experiment is predefined. Therefore, given the $n$ observations generated from the distribution under study, we define a time $t_c$ such that any component with a survival time exceeding this fixed time is considered a censored observation. To simulate this type of censoring, consider the following steps:
\textit{algorithm}.
\begin{algorithm}[H]
\caption{Simulation With Type I censorship}
\label{wsimultypeicensorship}
\begin{algorithmic}[1]
\STATE \mbox{Define} $n$, $t_c$ and $\boldsymbol{\theta}$, 
\FOR{$i=1$ to $n$}
\STATE $u_i \leftarrow U(0,1)$
\STATE $x_i \leftarrow F^{-1}(u_i,\boldsymbol\theta)$
\IF{$x_i<t_c$} 
\STATE $t_i\leftarrow x_i$ 
\STATE $\delta_i \leftarrow 1$
\ELSE 
\STATE $t_i\leftarrow t_c$ 
\STATE $\delta_i \leftarrow 0$
\ENDIF
\ENDFOR
\PRINT $t$, $\delta$
\end{algorithmic}
\end{algorithm}

As can be noted, we defined $\delta$ as the indicator of censoring, i.e.,  if $\delta=1$, the component has observed the event of interest, whereas $\delta=0$ indicates that the component is censored.

\begin{example}\label{exctype1}

Let us consider again that $X\sim\text{Weibull}(\beta,\alpha)$, to simulate data with type I censorship. We set the parameter values and the time that the experiment $t_c$  will end. Assuming that $\alpha=1.5$, $\beta=2.5$ and $t_c=1.5$ the codes to generate the censored data are given as follows.
\end{example} 
\begin{lstlisting}[language=R]
#####Weibull R data simulation With Type I 
n=50  #Sample size
tc<- 0.51 ##Censored time
beta <- 2.5 ##Parameter value
alpha <- 1.5 ##Parameter value
t<-rweibull(n, alpha, beta)
delta<-rep(1,n) 
for (i in 1:n)
  if(t[i]>=tc){
  t[i]<-tc
  delta[i]<-0 }
\end{lstlisting}

The obtained output in R is given by

\begin{verbatim}
> print(t)
 [1] 0.51 0.34 0.51 0.41 0.15 0.09 0.15 0.34 0.01 0.51 0.51 0.51 0.51 0.36 0.35
[16] 0.46 0.51 0.51 0.46 0.25 0.20 0.10 0.51 0.51 0.51 0.17 0.31 0.51 0.39 0.47
[31] 0.51 0.10 0.51 0.51 0.01 0.32 0.03 0.51 0.44 0.22 0.51 0.22 0.51 0.51 0.51
[46] 0.49 0.40 0.51 0.51 0.17
> print(delta)
 [1] 0 1 0 1 1 1 1 1 1 0 0 0 0 1 1 1 0 0 1 1 1 1 0 0 0 1 1 0 1 1 0 1 0 0 1 1 1 0
[39] 1 1 0 1 0 0 0 1 1 0 0 1

\end{verbatim}
 
An important concern is how we set $t_c $ to obtain a desirable level of censored data. For instance, what is the ideal $t_c$ to obtain $20\%$ of censored observations? Although in applications, the $t_c$ is usually defined, here, we have to find the value of $t_c$ to generate the pseudo-random samples. To achieve that, we can consider the inverse transformed method since, 
\begin{equation*}
    F(t_c;\boldsymbol{\theta})=1-\pi \quad \Rightarrow \quad  t_c=F^{-1}(\pi; \boldsymbol{\theta}).
\end{equation*}

Hence, if we want $40\%$ censorship, the $t_c$ should be found in the quantile function at point $0.6$, i.e.,
\begin{equation*}
    \int_{0}^{t_c}f_{X}(x)dx=1-\pi,
\end{equation*}
where $\pi$ is the percentage of censorship. Then, if $x> t_c$ is censorship $x \leq t_c$ is complete then we expect $0.6$ to be complete and $0.4$ censored. For example, in the case where the data is generated from the Weibull distribution, we will have the following,
\begin{equation}
    \int_{0}^{t_c} \alpha \beta x^{\alpha-1}\exp\{-\beta x^{\alpha}\}dx=1-\pi.
\end{equation} 
The quantile function for our scheme takes the form
\begin{equation*}
 t_c=\left(\frac{\left(-\log(\pi)\right)}{\beta}\right)^{\frac{1}{\alpha}}.
\end{equation*} 
Hence, for values of $\pi = 0.4$, $\beta = 2.5$, and $\alpha = 1.5$, we find that $t_c = 0.5121$. In fact, running the code of Example \ref{exctype1} with $n = 50000$, we obtain $\hat{\pi} = 0.3969$.

\subsection{Random censoring}

In the simulation method with random censorship (type III) for survival data, two distributions should be generated. The first distribution governs the survival times in the absence of censorship while the second distribution dictates the censorship mechanism. In this context, a censored event occurs when the uncensored survival time, generated by the first distribution, exceeds the censored time produced by the second distribution.

Specifically, let us consider a non-negative random variable $X$, representing our first distribution that generates the times until an event occurs. The survival function $\mathcal{S}_{X}(x)$, derived from the cumulative distribution function $F_{X}(x)$, provides the probability of survival beyond time $x$. This is expressed as $F_{X}(x)=P(X\leq x)$ and $\mathcal{S}_{X}(x)=1-P(X\leq~x)=P(X>x)$, where $f_{X}(x)$ denotes the probability density function of $X$. For the second distribution, which controls the censorship mechanism $C$, the selection is at the researcher's discretion. However, employing a continuous distribution is advisable. In this example, we assume that the random variable $X$ follows the $\text{Weibull}(\beta, \alpha)$ distribution and that $C$ is uniformly distributed, $C \sim U(0,\lambda)$. We further assume independence between these two distributions $(X\perp U)$. Thus, our censored samples adhere to the following scheme: for each $x$-th observation, we record $(T_x, \delta_x)$, where $T_x = \min(X_x, C_x)$ and $\delta_x = \mathrm{I}(X_x \leq C_x)$. Here, $\delta = 1$ indicates an observed lifetime, whereas $\delta = 0$ signifies a censored one. The subsequent algorithm details this random censoring process.

\begin{algorithm}[H]
\caption{Simulation With Random censoring}
\label{wsimultyper1censorship}
\begin{algorithmic}[1]
\STATE \mbox{Define} $n$, $\lambda$ and $\boldsymbol{\theta}$, 
\FOR{$i=1$ to $n$}
\STATE $u_i \leftarrow U(0,1)$
\STATE $x_i \leftarrow F^{-1}(u_i,\boldsymbol\theta)$
\STATE $c_i \leftarrow U(0,\lambda)$
\IF{$x_i<c_i$} 
\STATE $t_i\leftarrow x_i$ 
\STATE $\delta_i \leftarrow 1$
\ELSE 
\STATE $t_i\leftarrow c_i$ 
\STATE $\delta_i \leftarrow 0$
\ENDIF
\ENDFOR
\PRINT $t$, $\delta$
\end{algorithmic}
\end{algorithm}


\begin{example}\label{exeran}
Assuming that the distribution that controls censorship $C\sim U(0,\lambda)$, we generate data with random censorship for the time generated from the $X\sim\text{Weibull}(\beta,\alpha)$ distribution. The code in R is described as follows.
\end{example}

\begin{lstlisting}[language=R]
n=50;  beta <- 2.5; alpha <- 1.5 #Define the parameters
lambda<-1.21
t<-c()
y<-rweibull(n, alpha, beta)
delta<-rep(1,n) 
c<-runif(n,0,lambda)
for (i in 1:n) {
	if (y[i]<=c[i]) {
		t[i]<-y[i];
		delta[i]<-1
		}
	else {
		t[i]<-c[i];
		delta[i]<-0
		} 
	}
\end{lstlisting}

An example of output is given as follows.
\begin{verbatim}
> print(t)  
 [1] 0.56 0.34 0.53 0.41 0.15 0.00 0.15 0.27 0.01 0.53 0.25 0.67 0.28 0.36 0.35
[16] 0.46 1.19 0.56 0.46 0.20 0.20 0.10 0.47 0.57 0.65 0.17 0.31 0.07 0.39 0.47
[31] 1.00 0.10 0.37 0.48 0.01 0.21 0.03 0.44 0.15 0.22 0.73 0.22 0.43 0.91 0.54
[46] 0.22 0.40 0.19 0.74 0.17
> print(delta)
 [1] 1 1 1 1 1 0 1 0 1 1 0 1 0 1 1 1 1 1 1 0 1 1 0 0 0 1 1 0 1 1 0 1 0 0 1 0 1 0
[39] 0 1 0 1 0 1 1 0 1 0 1 1
\end{verbatim}

To find the desired censorship percentage in the random censorship scheme, we assumed independence between the random variable $C_x \sim F_C(\cdot \mid \lambda)$ and $X_x \sim F_X(\cdot \mid \theta)$ (continuous distributions), in which $\pi$ was the desired percentage of censorship. Note that in the case of censored data, we have the following situation, $(C_x,0) \iff T_x = C_x \iff C_x < X_x$. Hence, by definition, $P(C_x<X_x)$, for $C_x \ \perp \ X_x$, is given by
\begin{equation}\label{Prob}
\begin{aligned}
    \pi &= P(C_x < X_x) \\ &= \int_0^\infty P(C_x< x \mid X_x = x) f_X(x) dx \\
    &=  \int_0^\infty F_C(x\mid \lambda) f_X(x) dx \\ 
   \pi &= \mathbb{E}_X[F_C(X\mid \lambda)],     
\end{aligned}
\end{equation}
we propose to find the solution of the parameter $\lambda$ of the previous equation by approximating the expectation by the law of large numbers, in the following way:
\begin{equation}\label{MC}
\begin{aligned}
    \pi &= \mathbb{E}_X[F_C(X\mid \lambda)]  \\
    \pi &\approx \dfrac{1}{n} \sum_{i=1}^n F_C(x_i\mid \lambda), \ \ \text{with } x_i\sim F_X(\cdot \mid \theta).
\end{aligned}
\end{equation}

For instance, recalling Example 4.3, assuming that the same parameter values $\alpha = 1.5$ and $\beta = 2.5$ but setting $\pi = 0.4$, we have that
$\lambda = 1.206$.
\\
The algorithm and the code in R for this example is given as follows.

\begin{algorithm}[H]

\caption{Find parameter to desired censorship percentage}
\label{Find_Parameter}
\begin{algorithmic}[1]
\STATE Define  $\boldsymbol n,\ \boldsymbol \theta$ \ \text {and} $\boldsymbol\pi$
\FOR{$i=1$ to $ n$}
\STATE $u_i \leftarrow U(0,1)$
\STATE $x_i \leftarrow F_X^{-1}(u_i\mid \boldsymbol\theta)$
\ENDFOR
\STATE $\boldsymbol\lambda_{mc}  \gets $
    solve$\displaystyle \{\boldsymbol\pi  - \dfrac{1}{n} \sum_{i=1}^n F_C(x_i\mid \boldsymbol\lambda) = 0 \}$
\PRINT $\boldsymbol\lambda_{mc} $
\end{algorithmic}

\end{algorithm}

\begin{example}\label{exerFind}
 We determine the parameter for the desired censorship percentage. Assuming that the distribution controlling censorship is $C \sim U(0, \lambda)$, we generate data with random censorship for the time generated from the $X \sim \text{Weibull}(\beta, \alpha)$ distribution. The code in R is described as follows:
\end{example}

\begin{lstlisting}[language=R]
beta <- 2.5; alpha <- 1.5 #Define the parameters distribution
pi <- 0.5 # Define censorship percentage
n_mc <- 1e06; e_min <- 1e-04; e_max  <- 1e06  # MC parameters

x <- rweibull(n_mc, alpha, beta)
f <- function(lambda){mean(punif(x, min = 0, max = lambda))- pi}
# Find solution f(lambda) = 0
lambda_mc <- uniroot(f, interval = c(e_min, e_max))$root 

\end{lstlisting}

An example of output is given as follows.
\begin{verbatim}
> print(lambda_mc)  
 [1] 1.204695
\end{verbatim}

To generate left-censored data, we simply use the same procedure for right-censored data, but instead apply the censorship left tail of the distribution. $(C_x, 0) \Leftrightarrow T_x = C_x \Leftrightarrow X < C_x $
Furthermore, the relationship between the censoring probability and the left-censored data can be expressed as:

\begin{equation}\label{left-censored}
\begin{aligned}
\pi = P(X < C_x) \ \iff \
1 - \pi = P(C_x \leq X)
\end{aligned}
\end{equation}

Note that in our approach, a large sample size is considered. This is because, as the sample size increases, the percentage of censoring tends to reflect the true value more accurately. If a large number of samples are collected and their average is taken, it is expected to converge to a mean of $\pi$. A significant limitation of this method, however, is that the parameter estimation is performed using the Monte Carlo method. This technique provides an approximation in probability of the desired value, rather than a precise calculation.

\subsection{Sampling with cure fraction}

In medical and public health studies, it is common to encounter a proportion of individuals who are not susceptible to the event of interest. These individuals are often referred to as immune or cured, and the fraction of such individuals is known as the cure fraction. Survival models that incorporate this feature are known as cure rate models or long-term survival models \citep{maller1996survival}.

\cite{rodrigues2009unification} proposed a unified approach to these long-term survival models. In this framework, let $M$ represent the number of competing causes related to the occurrence of an event of interest. Assume that $M$ follows a discrete distribution with a probability mass function (pmf) denoted by $p_m=P(M=m)$, for $m=0,1,2,\ldots$. Given $M=m$, consider $Y_1,Y_2,\ldots,Y_m$ as independent identically distributed (iid) continuous non-negative random variables that are independent of $M$ and share a common survival function $S(y)=P(Y\geq y)$. The random variables $Y_k$, for $k=1,2,\ldots,m$, represent the time-to-failure of an individual due to the $k$-th competing cause. Consequently, the observable lifetime is defined as $X=\min\{Y_1,Y_2,\ldots,Y_M\}$ for $M\geq 1$, with $P(Y_0=+\infty)=1$, indicating a proportion $p_0$ of the population that is not susceptible to failure. The long-term survival function of the random variable $X$ is therefore given by
\begin{eqnarray}\label{crm}
S_{\rm pop}(x)&=&P\left(X\geq x\right)\nonumber\\
&=&P\left(M=0\right)+P\left(Y_1>x,Y_2>x,\ldots,Y_M>x,M\geq 1\right)\nonumber\\
&=&p_0+\sum_{m=1}^{\infty}p_m P\left(Y_1>x,Y_2>x,\ldots,Y_M>x \mid M=m\right)\nonumber\\
&=&\sum_{m=0}^{\infty}p_m\left[S(x)\right]^{m}\nonumber\\
&=&G_M\left(S(x)\right),
\end{eqnarray}
where $S(x)$ is a proper survival function associated with the individuals at risk in the population and $G_M\left(\cdot\right)$ is the probability generating function (PGF) of the random variable $M$, whose sum is guaranteed to converge because $S(t)\in[0,1]$  \citep{feller2008introduction}. 

Suppose that the number of competing causes follows a Binomial negative distribution in equation \eqref{crm}, with parameters $\kappa$ and $\gamma$, for $\gamma>0$ and $\kappa\gamma\geq -1$, following the parameterization of \cite{piegorsch1990maximum}. In this case, the long-term survival model is given by
\begin{eqnarray}\label{bin}
S_{pop}(x)= \left(1+\kappa\gamma(1-S(x))\right)^{-1/\kappa}, \quad x>0.
\end{eqnarray}

The Negative Binomial distribution includes the Bernoulli ($\kappa=-1$ and $m=1$) and Poisson ($\kappa=0$) distributions as particular cases. Hence, equation \eqref{bin} reduces to the mixture cure rate model, proposed by \cite{berkson1952survival}, and to the promotion time cure model, introduced by \cite{tsodikov1996stochastic}, respectively. Here, we focus on the mixture cure rate model as most of the cure rate models can be written as
\begin{eqnarray}\label{bin2}
S_{pop}(x)= p+(1-p)S(x), \quad x>0 \quad \mbox{and} \quad 0\leq p\leq 1.
\end{eqnarray}

An algorithm to generate a random sample for the model above is shown as follows.
\begin{algorithm}[H]
\caption{Sampling With Cure fraction}
\label{Wsamplingcurefration}
\begin{algorithmic}[1]
\STATE \mbox{Define} $n$, $\beta$, $\alpha$, $p$ and $\lambda$,
\FOR{$i=1$ to $n$}
\STATE $b_i \leftarrow \text{Ber}(n,p)$
\IF{$b_i=1$} 
\STATE $x_i\leftarrow \infty$ 
\ELSE 
\STATE $u_i \leftarrow U(0,1)$
\STATE $x_i \leftarrow F^{-1}(u_i,\boldsymbol\theta)$
\ENDIF
\ENDFOR
\FOR{$i=1$ to $n$}
\STATE $c_i \leftarrow U(0,\lambda)$
\IF{$x_i<c_i$} 
\STATE $t_i\leftarrow x_i$ 
\STATE $\delta_i \leftarrow 1$
\ELSE 
\STATE $t_i\leftarrow c_i$ 
\STATE $\delta_i \leftarrow 0$
\ENDIF
\ENDFOR
\PRINT $t$, $\delta$
\end{algorithmic}
\end{algorithm}

\begin{example}
Let $X\sim\text{Weibull}(\beta,\alpha)$ and suppose that the population is composed of two groups: (i) cured, with a cure fraction $p$, and (ii) non-cured, with a cure fraction $1-p$ and survival function $S_0(x;\alpha,\beta)=\exp\{-\beta x^{\alpha}\}$. Thus, the survival function of the entire population, $S_p(x;\alpha,\beta)$, is given by:
 \begin{equation}\label{scurefrac1}
 \begin{aligned}
 S_p(x;\alpha,\beta)&= p + (1-p)\exp\{-\beta x^{\alpha}\}.
 \end{aligned}
 \end{equation}

The pseudo-random sample with cure fraction using R can be generated using the following code.
\end{example}
\begin{lstlisting}[language=R]
n=50;  beta <- 2.5; alpha <- 1.5; p<-0.3 #Define the parameters
lambda<-1.21

require(Rlab)
##Generation of pseudo-random samples with long-term survival
b<-rbern(n,p)
t<-c(); y<-c()
for (i in 1:n) {
    if (b[i]==1) {
        y[i]<-1.000e+54
    }  
    else{
        y[i]<-rweibull(1, alpha, beta)
    } 
}
##Generation of pseudo-random samples with random censoring
delta<-c()
cax<-runif(n,0,lambda)
for (i in 1:n) {
	if (y[i]<=cax[i]) {t[i]<-y[i] ;delta[i]<-1}
	if (y[i]>cax[i]) {t[i]<-cax[i] ;delta[i]<-0}
}
\end{lstlisting}

The obtained output in R is given by

\begin{verbatim}
> print(t)
 [1] 0.48 0.26 0.21 0.29 0.44 0.01 0.27 0.22 0.46 0.41 0.97 0.62 0.22 0.21 0.19
[16] 0.22 0.52 0.37 0.32 0.53 1.14 0.31 0.20 0.83 0.96 0.56 0.18 0.67 0.04 0.52
[31] 1.13 0.11 0.49 0.35 0.40 0.45 0.27 1.19 0.26 0.08 1.02 0.07 1.09 1.22 0.63
[46] 0.77 0.12 0.79 0.63 0.28
> print(delta)
 [1] 0 1 0 0 0 1 0 1 1 1 0 0 0 1 0 1 0 1 0 1 0 0 0 0 0 1 1 0 0 0 0 0 0 0 1 1 1 0
[39] 1 1 0 0 0 0 1 1 1 0 0 1 1
\end{verbatim}

As can be seen in the results above, the proportion of censoring was $0.6$, which is different from $0.4$, which would be achieved by selecting $\lambda=1.21$. Since we are dealing with a new model with an additional parameter, the results need to be computed using the new PDF.
 Therefore, once this new PDF has been created, we need to obtain a random number generator from this distribution to call the Algorithm \ref{Find_Parameter}. Concerning another possibility to overcome this problem, we can consider a grid search approach to find the value of $\lambda$, as illustrated in the algorithm below.
\begin{algorithm}[H]
\caption{Finding the desirable censoring}
\label{Wcen1}
\begin{algorithmic}[1]
\STATE \mbox{Define} $n$, $\boldsymbol\pi$, $\pi_I$, $\lambda$,
\REPEAT
\STATE $\lambda= \lambda+\epsilon$
\STATE Call Algorithm 6
\STATE $\hat\pi=1-\frac{\sum_{i=1}^{n}(\delta_i)}{n}$

\UNTIL{$\pi_I\leq\hat\pi$}
\PRINT $\lambda$
\end{algorithmic}
\end{algorithm}

In the previously discussed algorithm, it is essential to establish an initial value for $\lambda$, the ideal proportion of censoring $\pi_I$, and to invoke Algorithm 6 during the iterative process. The search for the desired value of $\lambda$ is straightforward, but the computational time required may vary depending on the initial value of $\lambda$ and the chosen $\epsilon$, which is used to increment $\lambda$. A practical approach is to set $\lambda$ to the minimum value from a simulated sample, which often provides a better starting point than setting $\lambda$ to 0. The choice of $\epsilon$ should consider the data range; for most applications involving time, $\epsilon=0.01$ is a common choice. To achieve high precision, a large sample size is required — in practice, $n=10,000$ is typically used. It is crucial to note that while this approach has been presented in the context of long-term survival models, it is equally applicable to scenarios involving random censoring and type I censored data.

\begin{lstlisting}[language=R]
library(Rlab)
n=10000;  beta <- 2.5; alpha <- 1.5; p<-0.3 #Define the parameters
lambda<-min(rweibull(n, alpha, beta))
epsilon<-0.01
pi_i<-0.6 
pi<-1 #Initial value of pi

##Creting pseudo-random samples with long-term survival
while(pi_i<=pi) {
	lambda<-lambda+epsilon
	b<-rbern(n,p)
	t<-c(); y<-c()
	for (i in 1:n) {
		if (b[i]==1) {y[i]<-1.000e+54}  
		if (b[i]==0) {y[i]<-rweibull(1, alpha, beta)} 
	}
	##Creting pseudo-random samples with random censoring
	delta<-c()
	cax<-runif(n,0,lambda)
	for (i in 1:n) {
		if (y[i]<=cax[i]) {t[i]<-y[i] ;delta[i]<-1}
		if (y[i]>cax[i]) {t[i]<-cax[i] ;delta[i]<-0}
		}
	pi<-1-(sum(delta)/n)
}
\end{lstlisting}

The obtained output in R is given by

\begin{verbatim}
> print(pi)
 [1] 0.5938
> print(lambda)
 [1] 1.110444
\end{verbatim}


{\color{black}
\subsection{Package rcens}
}
All previously described algorithms \ref{wsimultypeii2censorship}, \ref{wsimultypeicensorship}, \ref{wsimultyper1censorship}, \ref{Find_Parameter} and \ref{Wsamplingcurefration} for simulating censored data type I, II, III and cure fraction are implemented in the new package ``\textbf{rcens}" available at CRAN in R. The functions of this package are implemented to be flexible to any desired distribution, only needing the sample generators of the distribution that will be censored $F_X$. In the case of random censoring, the sample generator of the distribution of censorship $F_C$. Based on this insight, we can generate censored data from any distribution. It also offers the option to generate left-censored data, as explained in equation (\ref{left-censored}).

These new generators of censored samples enable control over the level of censoring and cure fraction as desired, facilitating the performance and analysis of simulated experiments that aim to study the effects of modifying the level of censoring or cure fraction in their modeling.

In example \ref{exercensT3}, a practical use case is shown to simulate type 3 censoring. Where the function has the following arguments: \textit{rdistrX} sample generator of distribution to be censored, \textit{pdistrC} function distribution $F_C $, \textit{rdistrC} sample generator of censorship distribution, \textit{param\_X} list with the parameters of the distribution $F_X$, \textit{param\_C} list with the parameters of the distribution $ F_C$ where one of these must be the lambda parameter (which is adjusted to achieve the desired censoring), \textit{n} number of sample, \textit{theta}, desired censoring percentage and \textit{right} TRUE if it is desired to be right censoring or FALSE if left censoring is desired. More detailed information and usage examples for each function can be found in the package manual.

\begin{example}\label{exercensT3}
We create a sample censored data ($n=50$) with random censorship assuming that time generated is from the $X\sim\text{Exp}(2)$, the distribution that controls censorship $C\sim U(0,\lambda)$ and 
$40\%$ of censoring. 
The code with the package ``rcens" is described as follows.

\begin{lstlisting}[language=R]
library(rcens)
set.seed(0)
Data = rcensT3(rdistrX = rexp, pdistrC = punif, rdistrC = runif,
               param_X = list("rate" = 2),
               param_C = list("min" = 0, "max" = "lambda"),
               n = 50,  theta = .4, right = TRUE)
\end{lstlisting}
The obtained output in R is given by
\begin{verbatim}
Censorship percentage: 0.38
> print(Data$lambda)
[1] 1.11
> print(Data$sample_censored)
[1] 0.74 0.37 0.05 0.71 0.06 0.20 0.24 0.57 0.52 0.31 0.26 0.36 0.60 0.01
[15] 0.00 0.11 0.04 0.15 0.70 0.07 0.13 0.08 1.08 0.42 0.75 0.02 0.46 0.46
[29] 0.30 0.45 0.02 0.33 0.02 0.86 0.72  0.05 0.09 0.05 0.15 0.01 0.26 0.05
[43] 0.16 0.96 0.17 0.36 0.21 0.73 0.91 0.03

> print(Data$censored_indicator)
[1] 0 1 1 1 1 1 1 0 0 1 0 1 1 0 1 1 1 1 0 1 1 0 0 1 1 1 0 0 1 0 1 1 1 0 0 0 1 1 0 
[39]1 1 1 1 0 1 1 1 0 0 0
\end{verbatim}

\end{example}

\section{Simulation study}
In this section, we present a Monte Carlo simulation study to illustrate the application of the pseudo-random sample schemes discussed in previous sections. From these generated samples, we estimate the parameters using their respective MLEs. Let $\boldsymbol\theta$ represent the parameter vector of size $i=1,\ldots,k$, and let $\hat\theta_{i,1},\hat\theta_{i,2},\ldots,\hat\theta_{i,N}$ denote the corresponding MLEs. To evaluate the performance of the estimation method, we adopt the following criteria: average bias, mean squared error (MSE), and coverage probabilities (CPs), calculated respectively by
\begin{equation}\label{measures}
\text{Bias}_i=\dfrac{1}{N}\sum_{j=1}^{N}\left(\hat\theta_{i,j}-\theta_i\right), \quad \text{MSE}_i=\dfrac{1}{N}\sum_{j=1}^{N}\left(\hat\theta_{i,j}-\theta_i\right)^2  \ \ \text{and}  \ \ \text{CP}(\theta_i)=\dfrac{\sum_{j=1}^{N} I_{(a,b)}\left(\theta_i\right)}{N}, 
\end{equation}
where $I_{A}(x)$ is the indicator function, $a=\hat{\theta}_i - z_{\frac{\xi}{2}}\sqrt{H^{-1}_{ii}(\hat{\boldsymbol{\theta}})}$, and $b=\hat{\theta}_i + z_{\frac{\xi}{2}}\sqrt{H^{-1}_{ii}(\hat{\boldsymbol{\theta}})}$, with $z_{\xi/2}$where the $\left(\xi/2\right)$-th quantile is of a standard normal distribution. 

The Bias indicates the deviation of the estimated values' averages from the true parameter value, while the MSE quantifies the average squared errors. Estimators with Biases and MSEs closer to zero are generally preferred. As for the 95\% CP, for a large number of experiments and using a 95\% confidence level, the frequency of intervals encompassing the true values of $\theta_i$ should be close to 95\%. Although it is acknowledged that MLEs can be biased in small samples, this measure is considered as the Bias diminishes with the increasing sample size ($n$). Therefore, if the algorithm for generating pseudo-random samples with censoring is flawed, it would yield biased estimates even for large samples.

\subsection{Type-II, type-I and random censoring}

We consider the MLEs discussed in Appendix A without the additional long-term parameter. We generate N=100,000 random samples for each type-I, type-II, and random censored scheme and its respective maximum likelihood estimates to conduct such an analysis. The steps performed by the algorithm are the following:
\begin{algorithm}
\caption{Weibull with type-I,type-II, and random censoring}
\label{Weibul sem cura}
\begin{algorithmic}[1]

\STATE Define the parameter value of $\theta=(\alpha,\beta)$; 

\STATE Generate pseudo-random samples of size $n$ from Weibull distribution;

\STATE Generate the pseudo-random sample with censoring.

\STATE Maximize the log-likelihood function by using the MLE $\boldsymbol{\hat\theta}=(\hat\alpha,\hat\beta)$;

\STATE Repeat $N$ times steps 2 till 4.

\STATE Compute the measures presented in (\ref{measures}).

\end{algorithmic}
\end{algorithm}

\begin{figure}[!h]
	\centering
	\includegraphics[scale=0.64]{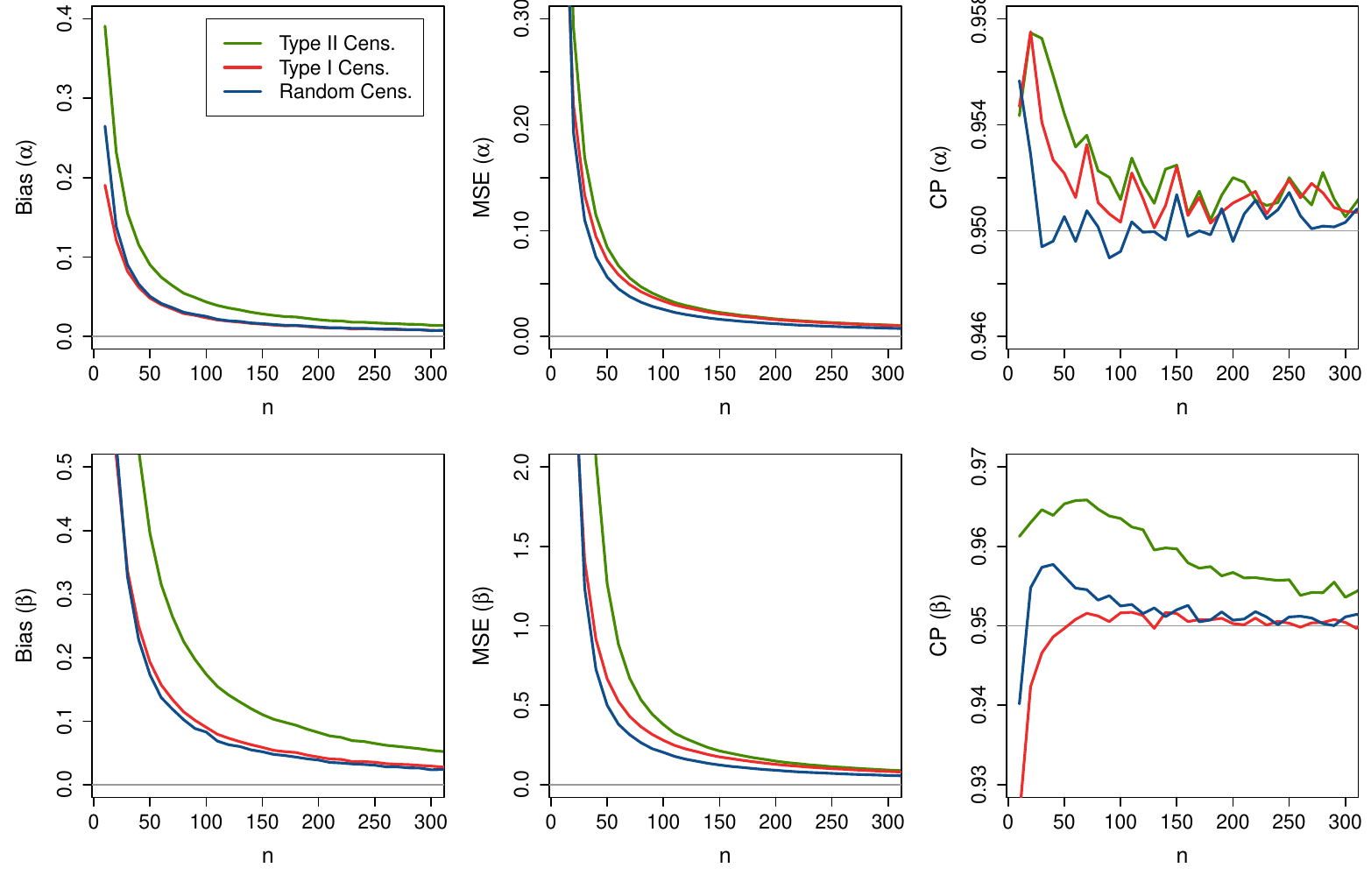}
	\caption{MREs, MSEs and CPs related to the estimates of $\alpha=2.5$ and $\beta=1.5$ for $N=100,000$ simulated samples, considering $n=(10,20,\ldots,300)$.}\label{fsimulation1}
\end{figure}

The parameter values are $\alpha=1.5$ and $\beta=2.5$, while the sample sizes are  $n\in\{10,40,60,\ldots,300\}$ with the proportion of censoring given by 0.4. Here, we used the same complete sample for the three different censoring schemes, such that we can observe the effect of censoring in the Bias and MSE. Figure \ref{fsimulation1} shows the Bias of the MSE and the coverage probability for the Weibull distribution parameters.

From the graph above, we can see that as $n$ increases, both bias and MSE are closer to zero, as expected. Moreover, the CPs are also closer to the $0.95$ nominal level. Thus, the proposed schemes to obtain censored samples do not introduce additional bias into the MLEs. Therefore, our proposed approaches can be further modified for different models.

\subsection{Long-term Weibull with random censoring}

The estimation assuming the presence of a long-survival term includes an additional parameter to be estimated. Hence, we have considered such an analysis in separate sections to account for this additional parameter. The algorithm \ref{Wsamplingcurefration} is used to obtain the pseudo-random samples with censoring. In contrast, the maximum likelihood estimators, discussed in Appendix A, are considered to obtain the estimates of parameter vector $\boldsymbol{\theta}=(\alpha,\beta,p)$ from the long-term Weibull distribution. The algorithm is described as follows:
\begin{algorithm}[H]
\caption{Long-term Weibull with random censoring}
\begin{algorithmic}[1]

\STATE Define the parameter value of $\theta=(\alpha,\beta,p)$; 

\STATE Generate a random sample of size $n$ from long-term Weibull distribution;

\STATE Obtain the censored data using Algorithm \ref{Wsamplingcurefration};

\STATE Maximize the log-likelihood function by using the MLEs presented in (\ref{mlecurefractionrandom});

\STATE Repeat $N$ times steps 2 until 4.

\STATE Compute the measures presented in (\ref{measures}).

\end{algorithmic}
\end{algorithm}

\begin{figure}[!h]
	\centering
	\includegraphics[scale=0.67]{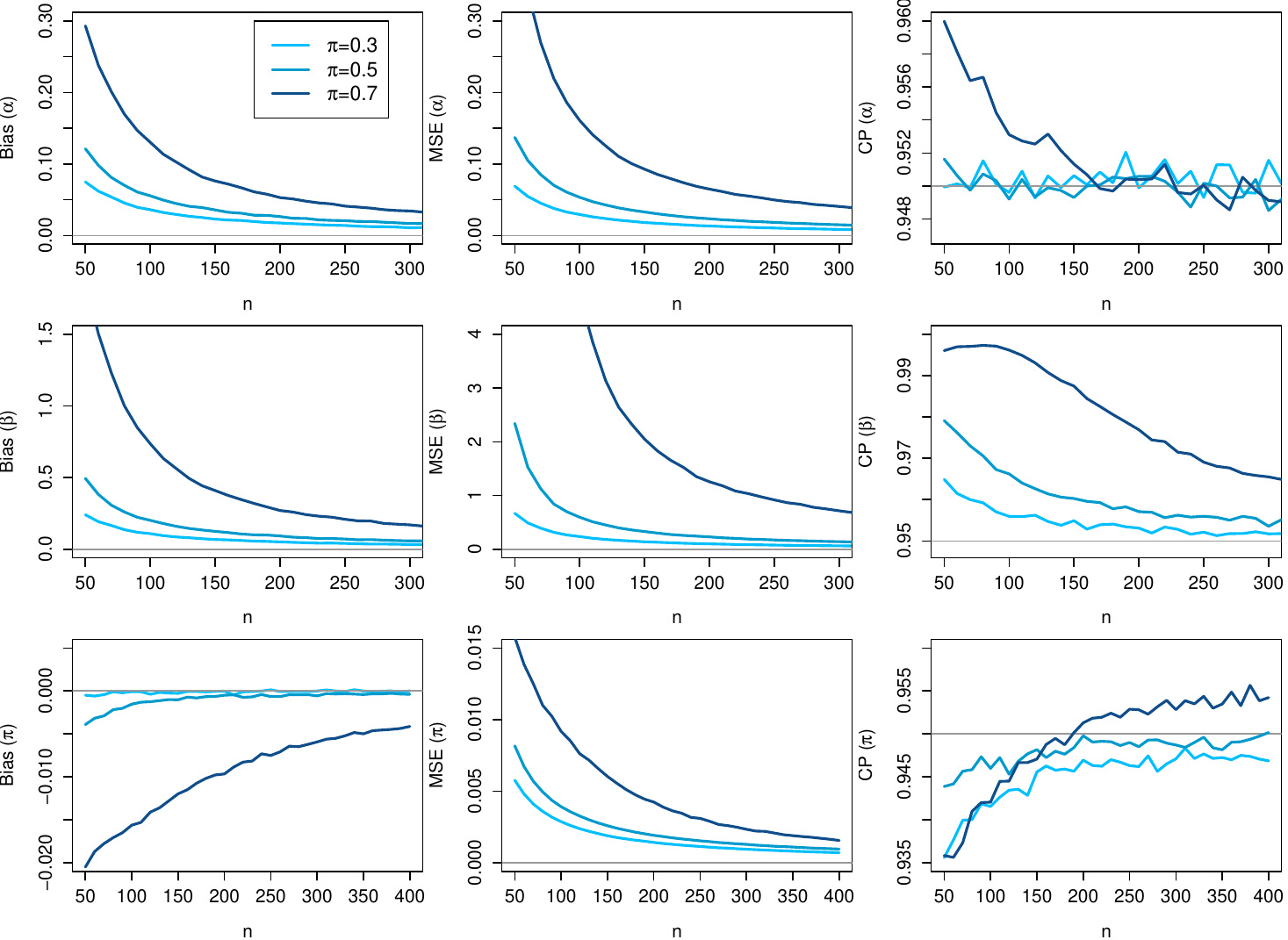}
	\caption{MREs, MSEs and CPs related to the estimates of $\alpha=1.5$, $\beta=2.5$ and $\pi=c(0.3,0.5,0.7)$ for $N=100,000$ simulated samples, considering different values of $n=(50,60, \ldots, 400)$.}\label{fsimulation2}
\end{figure}

Assuming the same steps as in the previous subsection, we compute the average bias, mean squared error (MSE), and coverage probability (CP) for each $\theta_i$, for $i=1,2,3$. We define  $n\in\{40,50,\ldots, 400\}$ and $N=100,000$, with  $\alpha=1.5$, $\beta=2.5$ and $\pi=(0.3,0.5,0.7)$. Hence, we expect 3 different levels of cure fraction in the population. In these cases, the censoring proportions are, respectively $0.4$, $0.6$, and $0.8$. To achieve such proportions, we have selected $\lambda=(3.110,2.159,1.389)$, obtained by equation (\ref{Wcen1}). Figure \ref{fsimulation2} displays the obtained Bias, MSE, and CP of the estimates of the long-term survival model and random censoring data. 

It can be seen from the results in Figure \ref{fsimulation2} that as $n$ increases, both bias and MSE go to zero, as expected. Additionally, the CPs also converge to the $0.95$ nominal level. The different levels of long-term survival and censoring indicate that as the proportion of censored data increases, uncertainty is included in the Bias, and the MSE of the estimates also increases. On the other hand, as we have large samples, the Bias tends to the true value regardless of the proportion of censored data.

\section{Conclusions}

Censored information occurs in nature, society, and engineering~\cite{klein2006survival}. Due to their relevance, several approaches were developed to consider such partial information during the statistical analysis. Despite the theoretical and applied advances in this area, the necessary mechanism to generate pseudo-random samples has not been considered in a unified framework. This paper has provided a unified discussion of the most common approaches to simulate pseudo-random samples with right-censored data. The censoring schemes considered were the most widely observed in real situations: type-I, type-II, and random censoring. The latter censoring scheme is general in the sense that the first two are particular cases. 

We presented a comprehensive discussion about the most common methods used to obtain pseudo-random samples from probability distributions, such as the inverse transformation method, an approach to simulate mixture models, and the Metropolis-Hasting algorithm, which is a Markov Chain Monte Carlo method. Furthermore, we provided a unified discussion on the necessary steps to modify the pseudo-random samples to include censored information. Additionally, we also discussed the necessary steps to achieve a desirable proportion of censoring. The proposed sampling schemes were illustrated using the Weibull distribution, one of the most common lifetime distributions. Furthermore, we also considered another probability distribution in which the sampling procedure and the estimation method are not easily implemented, i.e., the power-law distribution with cut-off. This distribution was considered, and the Metropolis-Hasting algorithm was used to obtain pseudo-random samples. The maximum likelihood estimators were also presented to validate the proposed approach. The codes in language R are presented and discussed to allow the readers to reproduce the procedures assuming other distributions.

The maximum likelihood method was used for estimating the parameters of the Weibull distribution by considering different types of censoring and long-term survival. We showed that the estimated parameters were closer to the true values. If the samples were not generated correctly, it would be expected that even for large samples, a systematic bias would be presented in the estimates, which were not observed during the simulation study. The coverage probabilities of asymptotic intervals also presented the expected behavior assuming that the censored information was generated correctly.

Our approach is general, and our results can be applied to most distributions introduced in the literature. These results also open up new opportunities to analyze censored data in models that are under development. Although we have not discussed a regression structure presence, the inverse transformed method can be used in the CDF with regression parameters. These results can also be applied in frailty models \citep{wienke2010frailty} since the unobserved heterogeneity is usually included using a probability distribution, which, combined with the baseline model, leads to a generalized distribution. Hence, both inverse transformed methods or Metropolis-Hasting algorithm can be used to sample complete data. Therefore, our approach can be used to modify the data with censoring. There are a large number of possible extensions of this current work. Other censoring mechanisms, such as left-truncated data and the different progressive censoring mechanisms, can be investigated using our approach;  this is an exciting topic for further research.

\section*{Disclosure statement}

No potential conflict of interest was reported by the author(s)

\section*{Acknowledgements}

Daniel C. F. Guzman and Alex L. Mota acknowledge the support of the coordination for the Improvement of Higher Education Personnel - Brazil (CAPES) - Finance Code 001.
Francisco Rodrigues acknowledges financial support from CNPq (grant number 309266/2019-0). Francisco Louzada is supported by the Brazilian agencies CNPq (grant number 301976/2017-1) and FAPESP (grant number 2013/07375-0). 

\section*{Code Availability}

\textcolor{black}{All the functions and procedures concerning implementation included in the article have been implemented in R Core Team. The codes are available as \url{https://github.com/dlsaavedra/rcens}.}

\section*{Apendix A - MLE for the Weibull distribution}\label{mleesqsimula}

Here, we revisit the MLEs for the Weibull distribution under different types of censoring and long-term survival. The results presented are not new in the literature and are standard in most statistical books related to survival analysis. The aim here is to estimate the distribution parameters using the pseudo-random samples presented in Section 4.

\subsection*{A.1 - Censorship type II}\label{sectionct2}

Let $X_1,\cdots,X_n$ be a random sample from Weibull$(\beta,\alpha)$. For type II censoring, the likelihood function is given by:
\begin{equation*}
L(\beta,\alpha;\boldsymbol{x})= \frac{n!}{(n-r)!} (\alpha\beta)^r \exp\left\{ -\beta \sum_{i=1}^{r} x_{(i)}^{\alpha}\right\} \left(\prod_{i=1}^{r} x_{(i)}\right)^{\alpha-1} \exp\left\{-\beta(n-r)x_{(r)}^{\alpha}\right\}.
\end{equation*}

The log-likelihood function without the constant term is given by:
\begin{equation*}
\ell(\beta,\alpha;\boldsymbol{x})\propto r\left[\log(\alpha)+\log(\beta)\right] - \beta \sum_{i=1}^{r} x_{(i)}^{\alpha} + (\alpha-1)\sum_{i=1}^{r} \log(x_{(i)}) - \beta (n-r)x_{(r)}^{\alpha}.
\end{equation*}

Setting $\dfrac{d}{d \beta}\ell(\beta,\alpha;\boldsymbol{x})=0$, $\dfrac{d}{d\alpha}\ell(\beta,\alpha;\boldsymbol{x})=0$ and after some algebraic manipulations, we obtain the MLE $\hat\beta$ of $\beta$,
\begin{equation*}
\hat\beta= \dfrac{r}{\sum_{i=1}^{r} x_{(i)}^{\hat\alpha}+(n-r)x_{(r)}^{\hat\alpha}},
\end{equation*}
where the MLE $\hat\alpha$ can be obtained as the solution of equation:
\begin{equation*}
\frac{r}{\alpha} +\sum_{i=1}^{r} \log(x_{(i)})= \left(\dfrac{r}{\sum_{i=1}^{r} x_{(i)}^{\alpha}+(n-r)x_{(r)}^{\alpha}}\right) \left[\sum_{i=1}^{r} x_{(i)}^{\alpha}\log(x_{(i)})+(n-r)x_{(r)}^{\alpha}\log(x_{(r)})\right].
\end{equation*}

The non-linear equation above has to be solved by numerical methods, such as Newton-Raphson, to find $\hat\alpha$. In R, we can use the next code to solve this equation.
\begin{lstlisting}[language=R]

likealpha<-function(alpha){
r/alpha+sum(log(w))-(r/(sum(w^alpha)+(n-r)*w[r]^alpha))*(sum(w^alpha*log(w))+
(n-r)*(w[r]^alpha)*log(w[r]))
}

###Simplifying the vector###
w<-t[1:r]

###Computing the MLE###
alpha<-uniroot(f=likealpha, c(0,100))$root
beta<-r/(sum(w^alpha)+(n-r)*w[r]^alpha)
\end{lstlisting}

To facilitate the computation, we have kept only the complete sample in the $w$ vector. The information related to the partial information is included directly by w[r]. The non-linear equation was included in a function that, combined with the uniroot function, can find the value of $\hat\alpha$. Using the same data generated in Example \ref{exctype2} and running the code above, the R output is given by
\begin{verbatim}
> print(c(alpha,beta))
[1] 1.25 1.92
\end{verbatim}

Hence, the obtained estimates are $\hat\alpha=1.25$ and $\hat\beta=1.92$.  These estimates differ when compared with the true value $\alpha=1.5$ and $\beta=2.5$. However, as the sample size increases, the values are closer to the true value. For instance, if we consider a larger sample of $n=50,000$ and $m=15000$, the estimates for the parameter will be $\hat\alpha= 1.49$ and $\hat\beta=2.50$.

\subsection*{A.2 - Censorship type I}

For type I censoring, we consider a random sample $X_1,\cdots,X_n$ from a Weibull distribution. In this case, the likelihood function is given by:
\begin{equation*}
L(\beta,\alpha;\boldsymbol{x})= \exp\left\{-\beta(n-r)x_{c}^{\alpha}\right\} \prod_{i=1}^{n} \left(\alpha\beta x_{i}^{\alpha-1} \exp\left\{-\beta x_{i}^{\alpha}\right\}\right)^{\delta_i}.
\end{equation*}

Then, the log-likelihood function can be expressed by:
\begin{equation*}
\ell(\beta,\alpha;\boldsymbol{x})=-\beta(n-r)x_{c}^{\alpha} + r\left[\log(\alpha)+\log(\beta)\right] + (\alpha-1)\sum_{i=1}^{n}\delta_i\log(x_i)-\beta\sum_{i=1}^{n}\delta_ix_{i}^{\alpha}.\nonumber
\end{equation*}

Considering $\dfrac{d}{d\beta}\ell(\beta,\alpha;\boldsymbol{x})=0$ and $\dfrac{d}{d\alpha}\ell(\beta,\alpha;\boldsymbol{x})=0$, we find the likelihood equations given by:
\begin{equation*}
 \frac{r}{\beta} =(n-r)x_c^{\alpha}+\sum_{i=1}^{n}\delta_ix_{i}^{\alpha}
\end{equation*}
and
\begin{equation*}
\frac{r}{\alpha} + \sum_{i=1}^{n}\delta_i\log(x_i)= \beta(n-r)x_{c}^{\alpha}\log(x_c) + \beta \sum_{i=1}^{n}\delta_ix_{i}^{\alpha} \log(x_i).
\end{equation*}
Hence, we obtain the MLE of $\beta$, $\hat\beta$ say, given by:

\begin{equation*}
\hat\beta= \dfrac{r}{(n-r)x_{c}^{\hat\alpha}+\sum_{i=1}^{n}\delta_i x_{i}^{\hat\alpha}},
\end{equation*}
where $\hat\alpha$ is the MLE of $\alpha$ obtained by solving the nonlinear equation:

\begin{equation*}
\frac{r}{\alpha} + \sum_{i=1}^{n}\delta_i \log(x_i) = \left( \dfrac{r}{(n-r)x_{c}^{\alpha}+\sum_{i=1}^{n}\delta_i x_{i}^{\alpha}}\right) \left[ (n-r)x_{c}^{\alpha}\log(x_c) + \sum_{i=1}^{n}\delta_i x_{i}^{\alpha}\log x_{i}\right]. \nonumber
\end{equation*}

The MLE for type I censoring has a similar structure as the one presented under type II censored data. Although $m=n-r$ is random here, the MLEs can be computed using similar techniques, as discussed in Section \ref{sectionct2}. The codes to solve the equation are presented as follows.

\begin{lstlisting}[language=R]
likealpha<-function(alpha){
r/alpha+sum(delta*log(t))-(r/(sum(delta*(t^alpha))+
(n-r)*tc^alpha))*((n-r)*(tc^alpha)*log(tc)+sum(delta*(t^alpha)*log(t)))
}

r<-sum(delta)
###Computing the MLE###
alpha<-try(uniroot(f=likealpha, c(0,100))$root)
beta<-r/(sum(delta*t^alpha)+(n-r)*tc^alpha)
\end{lstlisting}

Using the same data that were generated in Example \ref{exctype1} and running the code above, the output is given by
\begin{verbatim}
> print(c(alpha,beta))
[1] 1.26 1.96
\end{verbatim}

The estimates for the parameters are $\hat\alpha=1.26$ and $\hat\beta=1.96$. As in the previous example, as the sample size increases, the values become closer to $\alpha=1.5$ and $\beta=2.5$. For example, considering $n=50,000$ the estimates for the parameter would be $\hat\alpha= 1.50$ and $\hat\beta=2.52$ with $39.71\%$ of censored data.

\subsection*{A.3 - Random censorship}\label{randcensweib}

Assuming that $X_1,\cdots,X_n$ is a random sample from the Weibull distribution, the likelihood function considering data with random censoring can be written as:
\begin{equation*}
\begin{aligned}
L(\beta,\alpha;\boldsymbol{x}) = \prod_{i=1}^{n} \left(\alpha\beta x_{i}^{\alpha-1}\right)^{\delta_i} \exp\left\{-\beta x_{i}^{\alpha}\right\}.
\end{aligned}
\end{equation*}

The log-likelihood function is given by:
\begin{equation*}
\ell(\beta,\alpha;\boldsymbol{x})= r\left[\log(\alpha)+\log(\beta)\right] + (\alpha-1)\sum_{i=1}^{n}\delta_i\log(x_i) -\beta \sum_{i=1}^{n}x_{i}^{\alpha},
\end{equation*}
where $r=\sum_{i=1}^{n}\delta_i$ is the failure number ($r<n$). Thus, making $\dfrac{d}{d\beta}\ell(\beta,\alpha;\boldsymbol{x})=0$ and $\dfrac{d}{d\alpha}\ell(\beta,\alpha;\boldsymbol{x})=0$, the MLE of $\beta$ is given by:
\begin{equation*}
\hat\beta= \dfrac{r}{\sum_{i=1}^{n}x_{i}^{\hat\alpha}},
\end{equation*}
where $\hat\alpha$ is the MLE of $\alpha$, which can be determine by solving the nonlinear equation:
\begin{equation*}
\frac{r}{\alpha} + \sum_{i=1}^{n}\delta_i\log(x_i)=\left(\dfrac{r}{\sum_{i=1}^{n}x_{i}^{\alpha}}\right) \sum_{i=1}^{n}x_{i}^{\alpha}\log(x_i).
\end{equation*}

The MLEs for random censoring is a generalization of type I and II censoring, therefore, they can both be achieved using the equations above. Once more, the non-linear equation has to be solved and the function \textit{uniroot} is used to obtain the estimate of $\alpha$. The codes are given by:

\begin{lstlisting}[language=R]
likealpha<-function(alpha){
r/alpha+sum(delta*log(t))}- (r/(sum(t^alpha)))*sum(t^alpha*log(t))

###Simplifing some arguments###
r<-sum(delta)

###Computing the MLE###
alpha<-try(uniroot(f=likealpha, c(0,100))$root)
beta<-r/(sum(t^alpha))
\end{lstlisting}

The data presented in Example \ref{exeran} is considered here, from the code above the R output is given by
\begin{verbatim}
> print(c(alpha,beta))
[1] 1.34 2.10
\end{verbatim}

The estimates for the parameters, i.e., $\hat\alpha=1.34$ and $\hat\beta=2.10$, are not so far from the true values $\alpha=1.5$ and $\beta=2.5$. In fact, considering $n=50,000$, the estimates for the parameter are $\hat\alpha= 1.50$ and $\hat\beta=2.51$, with $39.44\%$ of censored data.

\subsection*{A.4 - Long-term survival with random censoring}\label{mlecurefractionrandom}

The presence of long-term survival includes an additional parameter. In this case, modifications in the PDF and the survival function are necessary. From the survival function (\ref{scurefrac1}), the long-term probability density function is given by 
\begin{equation*}
f(x;\beta,\alpha,p)=\alpha \beta\left(1-p \right)x^{\alpha-1}\exp\left(-\beta x^\alpha\right),
\end{equation*}
where $\beta>0$ and $\alpha>0$ are, respectively, the scale and shape parameters and $p\in (0,1)$ denotes the proportion of ``cured'' or ``immune'' individuals. Let $X_1,\ldots,X_n$ be a random sample of the long-term Weibull distribution, the likelihood function considering data with random censoring is given by:
\begin{equation*}
\begin{aligned}
L(\beta,\alpha,p;\boldsymbol{x, \delta})=\alpha^{r}\beta^{r}(1-p)^{r}\prod_{i=1}^{n}x_i^{\delta_i(\alpha-1)}\left[p+(1-p)\exp\left(-\beta x_i^\alpha\right)\right]^{1-\delta_i}\exp\left( -\beta\sum_{i=1}^{n}\delta_i x_i^{\alpha}\right),
\end{aligned}
\end{equation*}
where $r=\sum_{i=1}^{n}\delta_i$. Thus, the $\log$-likelihood function is:
\begin{equation}\label{eqlokcf2}
\begin{aligned}
\ell(\beta,\alpha,p;\boldsymbol{x, \delta})=&\ r\log(\alpha)+r\log(\beta)+r\log(1-p)+(\alpha-1)\sum^{n}_{i=1}\delta_{i}\log( x_i )-\beta\sum_{i=1}^{n}\delta_i x_i^{\alpha}\\ &+\sum_{i=1}^{n}(1-\delta_i)\log\left(p+(1-p)\exp\left(-\beta x_i^\alpha\right)\right).
\end{aligned}
\end{equation}

The likelihood equations are given by:
$$\frac{r}{\alpha}+\sum_{i=1}^{n}\delta_i\log(x_i)+\eta_1(\beta,\alpha,p;\boldsymbol{x, \delta})=\beta\sum_{i=1}^{n}\delta_ix_i^\alpha\log(x_i),$$

$$\frac{r}{\beta}+\eta_2(\beta,\alpha,p;\boldsymbol{x, \delta})=\sum_{i=1}^{n}\delta_ix_i^\alpha \quad \mbox{and} \quad \eta_3(\beta,\alpha;\boldsymbol{x, \delta})=\frac{r}{1-p},$$
where $(\theta_1,\theta_2,\theta_3)=(\beta,\alpha,p)$ and
\begin{equation*}
\eta_i(\beta,\alpha,p;\boldsymbol{x, \delta})= \sum_{i=1}^{n}(1-\delta_i)\dfrac{d}{d\theta_i}\log S(x_i,\beta,\alpha,p), \,\, \mbox{ for } i=1,2,3.
\end{equation*}

The solution of the likelihood equations can be achieved following the same approach discussed in the previous sections. On one hand, we can compute the estimates directly from the log-likelihood equation given by (\ref{eqlokcf2}). The codes necessary to compute the estimates are given as follows.
\begin{lstlisting}[language=R]
loglike<-function(theta) {
alpha<-theta[1]
beta<-theta[2]
p<-theta[3]
aux<- r*log(alpha)+r*log(1-p)+r*log(beta)+(alpha-1)*sum(delta*log(t))
+sum((1-delta)*log(p+((1-p)*(exp(-(beta*(t^alpha))))))) 
-sum(delta*(beta*(t^alpha)))
return(aux)
} 

###Simplifing some arguments###
r<-sum(delta)
###Initial Values###
alpha<-1.11
beta<-0.88
p<-0.6

library(maxLik)
##Calculing the MLEs
estimate  <- try(maxLik(loglike, start=c(alpha,beta,p)))
\end{lstlisting}

Note that we defined a function with the log-likelihood to find the maximum of the likelihood equation. We also used the package \textit{maxLik} \citep{henningsen2011maxlik}. The initial values need to be started to initiate the iterative procedure. Here, only for illustrative purposes do we use the MLEs of $\alpha$ and $\beta$ obtained from the random censoring that returned $\tilde\alpha=1.11$ and $\tilde\beta=0.88$, while we can consider for $p$ the proportion of censoring as the initial value $\tilde{p}=0.6$. The output of the previous code is given by
\begin{verbatim}
> coef(estimate)
[1] 1.705 4.023 0.444
\end{verbatim}

From the initial values, we observe that if the presence of long-term survival is ignored, the parameters estimates are highly biased, while the MLEs as $\hat\alpha=1.71$, $\hat\beta=4.02$ and $\hat{p}=0.44$ with $60\%$ of censored data. Although we have not discussed here, the maxLik function can consider different optimization routines, constraining the parameters, and setting the number of iterations \citep{henningsen2011maxlik}. Another important characteristic is the easy computation of the Hessian matrix and the standard errors used to construct confidence intervals. The matrix can be obtained from the argument \textit{hessian}, while the standard errors are obtained directly from \textit{stdEr} method. 
\begin{verbatim}
>print(estimate$hessian)
          [,1]       [,2]        [,3]
[1,] -21.11378  3.2862602  -10.238921
[2,]   3.28626 -0.8455459    3.772982
[3,] -10.23892  3.7729819 -127.069910

>print((-solve(estimate$hessian))^0.5)
      [,1]  [,2]  [,3]
[1,] 0.350 0.706 0.070
[2,] 0.706 1.841 0.246
[3,] 0.070 0.246 0.096

> stdEr(estimate)
[1] 0.350 1.841 0.096

\end{verbatim}

The standard errors for the estimated parameters under random censoring and long-term survival are given above and can be used to construct the asymptotic confidence intervals discussed in Section 2 and the coverage probabilities.


\section*{Appendix B - MLE for the power-law with cut off}

Here, we derived the maximum likelihood estimators for the parameter of the PLC distribution. Let $T_1,\ldots,T_n$ be a random sample such that $T$ has PDF given in (\ref{pdfplc}), then the likelihood function is given by
\begin{equation}\label{likeli1}
L(\alpha,\beta; \boldsymbol{x},x_{\min})=\frac{\beta^{n-n\alpha}}{\Gamma(1-\alpha,\beta x_{\min})^n}\prod_{i=1}^{n}x_i^{-\alpha}\exp\left(-\beta\sum_{i=1}^{n} x_i\right).
\end{equation}

The log-likelihood function $l(\alpha,\beta;\boldsymbol{t})=\log{L(\alpha,\beta; \boldsymbol{x},x_{\min})}$ is given by
\begin{equation}\label{loglikelihood}
\begin{aligned}
l(\alpha,\beta; \boldsymbol{x},x_{\min})=&n(1-\alpha)\log(\beta)-\alpha\sum_{i=1}^{n}\log(x_i)-\beta\sum_{i=1}^{n}x_i -n\log\Gamma\left(1-\alpha,\beta x_{\min}\right),
\end{aligned}
\end{equation}
and the score functions are given by
\begin{equation}\label{scoremle} 
\begin{aligned}
U(\alpha;\boldsymbol{x},x_{\min})=&-n\log(\beta)+\frac{G_{2,3}^{\,3,0}\!\left(\left.{\begin{matrix}1, 1\\0, 0 ,1-\alpha\end{matrix}}\;\right|\,\beta x_{\min}\right)}{\Gamma(1-\alpha,\beta x_{\min})} +\log(\beta x_{\min})-\sum_{i=1}^{n}\log(x_i)
\end{aligned}
\end{equation}
\begin{equation}\label{scoremle2} 
U(\beta;\boldsymbol{x},x_{\min})=-\frac{n(1-\alpha)}{\beta}-\sum_{i=1}^{n}x_i+\frac{e^{-\beta x_{\min}}}{\beta E_{\alpha}(\beta x_{\min})}.
\end{equation}

The maximum likelihood estimate is asymptotically normal distributed with a bivariate normal distribution given by
$ (\hat\alpha,\hat\beta) \sim N_2\left((\alpha,\beta),I^{-1}(\alpha,\beta)\right) \mbox{ for } n \to \infty$, where $I^{-1}(\alpha,\beta)$ is the inverse of the Fisher information matrix, where the elements are given by

\begin{equation*}
I_{\alpha,\alpha}(\alpha,\beta)=\frac{2n\Gamma(1-\alpha,\beta x_{\min})G_{3,4}^{\,4,0}\!\left(\left.{\begin{matrix}1, 1, 1\\0, 0, 0 ,1-\alpha\end{matrix}}\;\right|\,\beta x_{\min}\right)}{\Gamma(1-\alpha,\beta x_{\min})^2}
-n\frac{G_{2,3}^{\,3,0}\!\left(\left.{\begin{matrix}1, 1\\0, 0 ,1-\alpha\end{matrix}}\;\right|\,\beta x_{\min}\right)^2}{\Gamma(1-\alpha,\beta x_{\min})^2} ,
\end{equation*}
\begin{equation*}
I_{\alpha,\beta}(\beta,\alpha)=I_{\beta,\alpha,}(\alpha,\beta)=\frac{n}{\beta}-n x_{\min}e^{-\beta x_{\min}}\left[\frac{G_{2,3}^{\,3,0}\!\left(\left.{\begin{matrix}1-\alpha, 1-\alpha\\1-2\alpha, -\alpha ,-\alpha\end{matrix}}\;\right|\,\beta x_{\min}\right)}{\Gamma(1-\alpha,\beta x_{\min})^2}\right] ,
\end{equation*}
\begin{equation*}
\begin{aligned}
I_{\beta,\beta}(\alpha,\beta)=-\frac{ne^{-2\beta x_{\min}}\left(1-e^{\beta x_{\min}}\left(\alpha+\beta x_{\min}\right)E_{\alpha}\left(\beta x_{\min}\right) \right)}{\beta^2E_{\alpha}\left(\beta x_{\min}\right)^2}  -\frac{n\left(\alpha-1 \right)}{\beta^2}.
\end{aligned}
\end{equation*}

The score function and the elements of the Fisher information matrix depend on Meijer-G functions which are not easy to compute. On the other hand, we can obtain the estimates directly from the maximization of the loglikelihood function. Additionally, we observe that observed Fisher information is the same as the expected Fisher information matrix as they do not depend on $x$, hence we can construct the confidence intervals directly from the Hessian matrix that is computed from the maxLik function. The necessary codes to compute the results above are displayed in:
\begin{lstlisting}[language=R]
#The incomplete gamma function
ig <- function(x,a) {x^(a-1)*exp(-x)}
#The loglikelihood function
loglike <- function(theta){
  alpha  <- theta[1]
  beta <- theta[2]
  l<- n*(1-alpha)*log(beta)-alpha*sum(log(x))-beta*sum(x)
	-n*log(integrate(ig, lower = (beta*xm), upper = Inf,a=(1-alpha))$value)
  return(l) } 

#Computing the MLEs and the Hessian matrix
res  <- maxLik(loglike, start=c(0.5, 0.5)) 
\end{lstlisting}

As initial values, we have used $\tilde\alpha=0.5$ and $\tilde\beta=0.5$. The output containing the parameter estimates as well as their related variances are given by:
\begin{verbatim}
> coef(res)
[1] 1.655 0.281
> stdEr(res)
[1] 0.428 0.140
\end{verbatim}

From the results above, we have that $\hat\alpha=1.655$ and $\hat\beta=0.281$ with the respective standard errors $0.428$ and $0.140$. The confidence intervals are constructed from the asymptotic theory and are $(0.817; 2.492)$ for $\alpha$ and $(0.007; 0.555)$ for $\beta$.

\bibliographystyle{chicago}

\end{document}